\documentclass[%
 reprint,
 amsmath,amssymb,
 aps,prb,
 superscriptaddress,
 longbibliography,
]{revtex4-2}

\usepackage[titletoc,toc,title]{appendix}
\usepackage{float}
\usepackage{graphicx}
\usepackage{dcolumn}
\usepackage{bm}
\usepackage{hyperref}
\hypersetup{
    colorlinks=true,
    linkcolor=blue,
    citecolor=blue,      
    urlcolor=blue
    }
\usepackage[capitalize]{cleveref}
\usepackage{bbold}
\usepackage{physics}
\usepackage[normalem]{ulem}

\newcommand{\R}{\mathbb{R}}

\newcommand{\twomatrix}[4]{\begin{bmatrix} #1 & #2 \\ #3 & #4 \end{bmatrix}}

\newcommand{\set}[1]{\{#1\}}

\renewcommand{\a}{\alpha}
\renewcommand{\b}{\beta}

\newcommand{\G}{\Gamma}
\renewcommand{\d}{\delta}

\renewcommand{\l}{\lambda}

\newcommand{\m}{\mu}
\newcommand{\n}{\nu}

\renewcommand{\th}{\theta}

\newcommand{\w}{\omega}

\renewcommand{\appendix}{%
    \setcounter{table}{0}
    \renewcommand{\thetable}{A\arabic{table}}%
    \setcounter{figure}{0}
    \renewcommand{\thefigure}{A\arabic{figure}}%
    \setcounter{equation}{0}
    \renewcommand{\theequation}{A\arabic{equation}}%
    \setcounter{subsection}{0}
    \renewcommand{\thesubsection}{A\arabic{subsection}}%
    \renewcommand{\thesection}{A}%
 }

\renewcommand{\vec}[1]{\bm{#1}}
\newcommand{\vecl}[2]{\vec{#1}^{(#2)}}

\newcommand{\vecll}[1]{\vec{#1}^{(\ell)}}

\begin{document}

\preprint{APS/123-QED}

\title{Low-energy moir\'e phonons in twisted bilayer van der Waals heterostructures}

\author{Jonathan Z. Lu}
\email{jlu@college.harvard.edu}
\affiliation{Department of Physics, Harvard University, Cambridge, Massachusetts 02138, USA}
\author{Ziyan Zhu}
\altaffiliation{Present address: Stanford Institute for Materials and Energy Sciences,
SLAC National Accelerator Laboratory, Menlo Park, CA 94025, USA}
\email{ziyanzhu@stanford.edu}
\affiliation{Department of Physics, Harvard University, Cambridge, Massachusetts 02138, USA}
\author{Mattia Angeli}
\affiliation{John A. Paulson School of Engineering and Applied Sciences, Harvard University, Cambridge, Massachusetts 02138, USA}
\author{Daniel T. Larson}
\affiliation{Department of Physics, Harvard University, Cambridge, Massachusetts 02138, USA}
\author{Efthimios Kaxiras}
\affiliation{Department of Physics, Harvard University, Cambridge, Massachusetts 02138, USA}
\affiliation{John A. Paulson School of Engineering and Applied Sciences, Harvard University, Cambridge, Massachusetts 02138, USA}

\date{\today}

\begin{abstract}
We develop a low-energy continuum model for phonons in twisted moiré bilayers, based on a configuration-space approach.  In this approach,  interatomic force constants are obtained from density functional theory (DFT) calculations of {\it untwisted} bilayers with various in-plane shifts. 
This allows for efficient computation of phonon properties for any small twist angle, while maintaining DFT-level accuracy. Based on this framework, we show how the low-energy phonon modes, including interlayer shearing and layer-breathing modes, vary with the twist angle. As the twist angle decreases, the frequencies of the low-energy modes are reordered and the atomic displacement fields corresponding to phonon eigenmodes break translational symmetry, developing periodicity on the moir\'e length scale. We demonstrate the capabilities of our model by calculating the phonon properties of three specific structures: bilayer graphene, bilayer molybdenum disulfide (MoS$_2$), and molybdenum diselenide-tungsten diselenide (MoSe$_2$-WSe$_2$). 
\end{abstract}

\maketitle

\section{\label{sec:intro}Introduction}
Two-dimensional (2D) van der Waals (vdW) multi-layer structures can be constructed by stacking 2D materials in a layer-by-layer fashion. If the stacking involves a twist of one layer by an angle $\theta$ with respect to the next layer, the relative
misorientation between atomic lattices produces a moir\'e interference pattern with a length scale larger than the lattice constants of the individual layers.
Such moir\'e bilayers can give rise to localized electronic states and tunable electronic properties \cite{carr2017twistronics,yankowitz2018dynamic,liu1903spin,song2019all,burg2019correlated,cao2020tunable,moriyama2019observation}. In twisted bilayer graphene, superconductivity and other unconventional correlated states
have been observed when the twist angle $\theta$ takes values near the ``magic angle"  \cite{cao2018unconventional,cao2018correlated,lu2019superconductors,xie2020nature}. Similar behavior has been shown in twisted bilayers composed of transition metal dichalcogenides (TMDCs), albeit often with decreased sensitivity to the twist angle~\cite{wang2019magic,li2019intrinsic,wu2018hubbard,liu2014evolution,wu2019topological,zhang2021spectroscopic,tran2019evidence,jin2019observation,alexeev2019resonantly,seyler2019signatures,rivera2015observation,van2014tailoring,angeli2020TMDs}. 
Determination of the role of phonons is a crucial step towards an understanding of these strongly correlated electronic states, particularly superconductivity~\cite{Wu_PRL, Lian_PRL, Cea_PNAS, Fabrizio_PRX2019}.
The collective vibration of a twisted bilayer can be strongly affected by the moir\'e pattern, forming the so-called moir\'e phonons~\cite{koshino2019moire,Fabrizio_PRX2019}. Moir\'e phonons are also of general experimental interest, and have recently been observed in graphene~\cite{gadelha2021localization}.
The variation of phonon properties with the twist angle allow Raman spectroscopy to be used to help characterize moir\'e bilayers~\cite{campos2013raman,wu2014resonant,huang2016low,lin2018moire,quan2021phonon}.

The interplay of two distinct length scales and the large number of atoms within a moir\'e cell make calculations of the phonon properties extremely challenging.
Due to the rapid scaling in computational cost with respect to the moir\'e length, direct calculations of phonons using density functional theory (DFT) are currently not feasible for twist angles less than a few degrees. Moreover, DFT calculations using periodic supercells can only be carried out for a specific set of commensurate angles. Instead, a continuum approximation has been successfully used to model the electronic structure and atomic relaxation of twisted bilayers and trilayers \cite{dai2016twisted,nam2017lattice,espanol2017discrete,espanol2018discrete,leconte2019relaxation,carr2018relaxation,triGr}. Continuum models have also been considered for moir\'e phonon analysis. The most efficient of these are empirical continuum models, but they leave out important details such as the out-of-plane degree of freedom and phonon modes that require coupling, for instance the shearing and layer-breathing modes \cite{koshino2019moire}. 

We propose a first-principles-based continuum model of moir\'e phonons, similar to the one developed by \citet{quan2021phonon} to help interpret the Raman spectra of twisted MoS$_2$ bilayers, but we have extended it to momenta beyond the $\Gamma$-point and have applied the model to several different bilayer systems. To bypass the need to compute the interatomic forces for all atoms in a large moir\'e supercell, we adopt a configuration space formalism to describe the local atomic environment. This approach has been used as the basis for a continuum model of atomic relaxations in twisted vdW heterostructures~\cite{cfg1,cfg2,cfg3,cfg4,carr2018relaxation}. 
Using this model we obtain the low-energy phonon properties of bilayer graphene, bilayer MoS$_2$, and a MoSe$_2$-WSe$_2$ heterostructure, and show how a relative twist between the layers modifies the phonon frequencies and eigenmodes. 

The paper is organized as follows. In Section~\ref{sec:csc} we derive the moir\'e dynamical matrix in configuration space, including twist-angle-dependent relaxation~\cite{carr2018relaxation}. We then compare our model predictions to the results of three other approaches in Section~\ref{sec:comparison}.
In Section~\ref{sec:results} we apply our model to three representative moir\'e bilayers and analyze their band structure and real space 
atomic displacements. Finally, we summarize our results and their implications in Section \ref{sec:summary}.

\section{\label{sec:csc}Configuration Space Continuum Model}
\begin{figure*}[ht!]
    \centering
    \includegraphics[width=\linewidth]{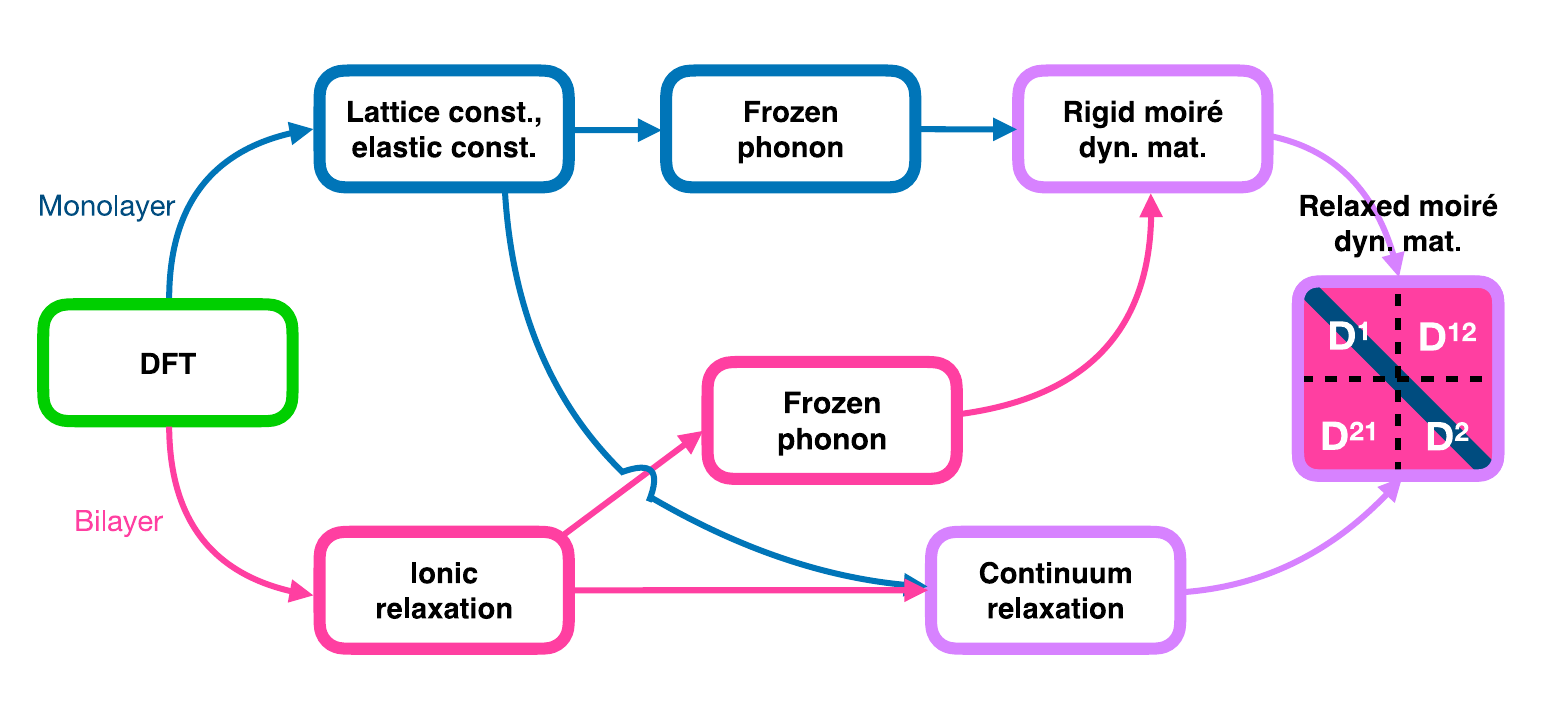}
    \caption{Schematic representation of the numerical implementation of the CSC model. A collection of monolayer (blue) and bilayer (pink) calculations are combined into a rigid (unrelaxed) moir\'e dynamical matrix and used to determine the bilayer relaxation pattern (both purple). All of these elements are combined to form the relaxed moir\'e dynamical matrix. The square labeled ``relaxed moir\'e dynamical matrix'' formally represents the dynamical matrix construction: the diagonal blocks are from the monolayer calculations whereas the off-diagonal blocks are constructed from the stacking dependent force field calculations.}
    \label{fig:workflow}
\end{figure*}

The workflow of our configuration space continuum (CSC) model is summarized in Fig.~\ref{fig:workflow}. We first construct the moir\'e dynamical matrix by computation of the dynamical matrices of many rigidly shifted bilayers---one for each configuration of atoms in layer 1 with respect to layer 2---and the individual monolayers. From the monolayer we obtain the optimized lattice constant and the elastic constants that are used for relaxation. We relax the ionic positions of each shifted bilayer as well, keeping the horizontal positions of one atom per layer fixed to maintain the registry between the layers. We then compute atomic forces using the frozen phonon approach, which under the CSC formalism produces the unrelaxed (rigid) moir\'e dynamical matrix. Finally, we include twist-dependent relaxation which yields the relaxed moir\'e form of the dynamical matrix. We provide a derivation of the model in this section, and details of the calculations can be found in Appendices~\ref{sec:GSFE} and~ \ref{sec:details}. Throughout the derivation, we use tildes to indicate a quantity in the moir\'e supercell, and when referring to the dynamical matrix we use a bar to denote the matrix in real space (the absence of the bar means the matrix is expressed in Fourier space).

\subsection{\label{sec:csc:sub:rigid}Moir\'e Dynamical Matrix}
The phonon spectrum of a crystal can be obtained by diagonalizing the dynamical matrix with the basis expanded in momentum space. 
The moir\'e dynamical matrix at a point $\widetilde{\vec{k}}$ 
in the moir\'e Brillouin zone, which we call the center site, is represented generically as a layer-block matrix encoding intralayer (diagonal) and interlayer (off-diagonal) couplings: \begin{align}
    D_{\text{moir\'e}}(\vec{\widetilde{k}}) = \begin{bmatrix}
    D^{1}(\vec{\widetilde{k}}) & D^{12} \\
    D^{12*} & D^2(\vec{\widetilde{k}}) .\label{eq:dmoire}
    \end{bmatrix} ,
\end{align}
Each block of the moir\'e dynamical matrix depends on two reciprocal-space degrees of freedom, $\vec{k}^{(\ell)}$ and $\vec{k}^{(\ell')}$, which are momenta from the monolayer Brillouin zones of layers $\ell$ and $\ell'$.
This is because in a twisted system, the interatomic forces depend not only on the pair-wise distance of two atoms, but also the position of the atom due to the change of the local environment.

\subsection{Moir\'e Scattering Selection Rule}
\label{sec:sub:selrule} 
We first derive how the degrees of freedom in the basis of Eq.~\eqref{eq:dmoire} are connected to each other through a selection rule. 
We begin with a monolayer, which is the intralayer term that has no stacking dependence. Considering only one layer $\ell$, the monolayer phonon equation of motion can be expressed as
\begin{align}
   \sum_{j\n\b} \overline{D}_{\m\n\a\b}^\ell(\vec{R}_i -  \vec{R}_j) \delta \widehat{u}_{j\ell\n\b}  = \omega^2 \delta \widehat{u}_{i\ell\m\a},
    \label{eq:intra-real-def}
\end{align}
where $\overline{D}_{\m\n\a\b}^\ell(\vec{R}_i -  \vec{R}_j)$ is the dynamical matrix element, $\vec{R}_i$ is a monolayer lattice vector, $\m,\n$ are Cartesian degrees of freedom, $\a,\b$ are sublattice degrees of freedom, and $\delta \widehat{u}_{i\ell\m\a}$ is the phonon displacement component that corresponds to atom $\a$ in direction $\m$. We can define a phonon displacement in the Bloch basis as follows, %
\begin{align}
\begin{aligned}
    \delta \widehat{u}_{i\ell \m \a}(\vecll{k}) & = \frac{1}{\sqrt{|\Gamma^*|}}e^{i\vec{k}^{(\ell)} \cdot \vec{R}_{i}} \delta u_{\ell\m\a}(\vec{k}^{(\ell)}),
\end{aligned}
\label{eq:intra-ft}
\end{align}
where $\vec{k}^{(\ell)}$
is a momentum degree of freedom in $\G^{(\ell)*}$, the monolayer Brillouin zone of layer $\ell$ with area $|\Gamma^*|$.  
We first note the orthogonality relation 
\begin{align}
\begin{aligned}
\sum_{\mu \alpha} \delta \widehat{u}_{i\ell\mu\alpha}^*  \delta \widehat{u}_{j\ell\mu\alpha} = \delta_{ij}.  \label{eq:ortho}
\end{aligned}
\end{align}
To expand the equation of motion, Eq.~\eqref{eq:intra-real-def}, in the Fourier basis, we multiply both sides by $\delta \widehat{u}_{k\ell \mu \alpha}^*$, then sum over $k$, $\mu$, and $\alpha$. 
\begin{align}
   \sum_{jk\m\n\a\b} \overline{D}_{\m\n\a\b}^\ell(\vec{R}_i -  \vec{R}_j) \delta \widehat{u}_{k\ell\mu\alpha}^*  \delta \widehat{u}_{j\ell\n\b}  &= \omega^2 \sum_{k\mu\alpha} \delta \widehat{u}_{k\ell\m\a}^* \delta \widehat{u}_{i\ell\mu\alpha}.
\end{align}
Using Eq.~\eqref{eq:ortho}:
\begin{align} 
   \sum_{ij\m\n\a\b} \overline{D}_{\m\n\a\b}^\ell(\vec{R}_i -  \vec{R}_j) \delta \widehat{u}_{i\ell\mu\alpha}^*  \delta \widehat{u}_{j\ell\n\b}  &= \omega^2.
\end{align} 
Relabeling $\vec{R} = \vec{R}_i, \Delta \vec{R} = \vec{R}_i - \vec{R}_j$ and 
changing to the Bloch basis using 
Eq.~\eqref{eq:intra-ft} yields
\begin{widetext}
\begin{align}
    \frac{1}{|\Gamma^*|} \sum_{\vec{R}} \left[ e^{- i (\vec{k}^{(\ell)} - \vec{k}'^{(\ell)}) \cdot \vec{R}} \sum_{\m\n\a\b} \sum_{\Delta \vec{R}} \overline{D}_{\m\n\a\b}(\Delta \vec{R}) e^{-i \vec{k}'^{(\ell)} \cdot \Delta \vec{R}} \delta u_{\ell\m\a}^*(\vec{k}^{(\ell)}) \delta u_{\ell\n\b}(\vec{k}'^{(\ell)}) \right] &= \omega^2.
    \label{eq:intra-rewrite}
\end{align}
\end{widetext}
Note the Poisson resummation rule: 
\begin{align}
	\sum_{\vec{R}} e^{ -i (\vecll{k}-\vec{k}'^{(\ell)}) \cdot \vec{R} } = |\Gamma^*| \sum_{\vecll{G}} \delta(\vecll{k}-\vec{k}'^{(\ell)}-\vecll{G}),\label{eq:sum}
\end{align} 
where $\vec{G}^{(\ell)}$ is a reciprocal lattice vector in layer $\ell$. 
Defining $D_{\m\n\a\b}(\vec{k}'^{(\ell)})$ as:
\begin{align}
    D_{\m\n\a\b}(\vec{k}'^{(\ell)}) = \sum_{\Delta \vec{R}} \overline{D}_{\m\n\a\b}(\Delta \vec{R}) e^{-i \vec{k}'^{(\ell)} \cdot \Delta \vec{R}},
\end{align} 
the phonon equation of motion becomes
\begin{align}
\begin{aligned}
    \sum_{\m\n\a\b} D_{\m\n\a\b}(\vec{k}^{(\ell)}) \delta u_{\ell\m\a}^*(\vec{k}^{(\ell)}) \delta u_{\ell\n\b}(\vec{k}^{(\ell)})  = \omega^2, 
\end{aligned}
\label{eq:intra}
\end{align}
for each $\vec{G}^{(\ell)}$, since the phonon displacement $\delta u_{i\ell\m\a}$ is the same at $\vecll{k}$ and $\vecll{k} + \vec{G}^{(\ell)}$. 
Equation~\eqref{eq:intra} is equivalent to: 
\begin{align}
\begin{aligned}
D(\vec{k}^{(\ell)} ) \cdot \delta \vec{u} (\vec{k}^{(\ell)})  = \omega^2 \delta \vec{u} (\vec{k}^{(\ell)}),\label{eqn:intra_eom} 
\end{aligned}
\end{align}
which is the eigenvalue equation that describes the intralayer phonon at $\vec{k}^{(\ell)}$, and the dynamical matrix $D(\vec{k}^{(\ell)})$ can be computed using either DFPT or frozen phonon approaches.

In a moir\'e cell the dynamical matrix gains an additional degree of freedom because the forces depend not only on the difference in lattice vectors, but also the real space position itself due to the twist. 
The general phonon equation of motion is given as the following:
\begin{align}
    \sum_{j\n\b\ell'} \overline{D}_{\m\n\a\b}(\vec{R}_i, \vec{R}_j)  \delta \widehat{u}_{j\ell'\n\b} = \omega^2 \delta \widehat{u}_{i\ell\m\a},
\end{align}
where $\vec{R}_i$ and $\vec{R}_j$ correspond to the atomic positions on layers $\ell$ and $\ell'$ respectively. 
Similar to the intralayer term, we multiply both sides of the equation by $\delta \widehat{u}^*_{i\ell\m\a}$ and sum over $\mu$, $\alpha$, and $\ell$. Plugging in the definition in Eq. \eqref{eq:intra-ft}:
\begin{widetext}
\begin{align}
     \sum_{\m\n\a\b\ell\ell'} D_{\widetilde{\vec{k}}\m\n\a\b}(\vec{k}^{(\ell)}, \vec{k}^{(\ell')}) \delta u^*_{\ell\m\a}(\vec{K}^{(\ell)}) \delta u_{\ell'\n\b}(\vec{K}^{(\ell')}) = \omega^2,
\end{align}
where 
%
\begin{align}
\begin{aligned}
    D_{\widetilde{\vec{k}}\m\n\a\b}(\vec{k}^{(\ell)}, \vec{k}^{(\ell')}) = \frac{1}{|\Gamma^*|} \sum_{ij} \overline{D}_{\m\n\a\b}(\vec{R}_{i}, \vec{R}_{j}) e^{-i \vec{K}^{(\ell)} \cdot \vec{R}_{i}} e^{i \vec{K}^{(\ell')} \cdot \vec{R}_{j}}.
\end{aligned}
\label{eq:inter-doubleFT}
\end{align}
\end{widetext}
and $\vecll{K} = \widetilde{\vec{k}} + \vecll{k}$ is an expansion for small momentum $\vecll{k}$ about a given point $\widetilde{\vec{k}}$ in the moir\'e Brillouin zone.
Making a two-center approximation such that $\overline{D}_{\m\n\a\b}(\vec{R}_{i}, \vec{R}_{j})=\overline{D}_{\m\n\a\b}(\vec{R}_{i}-\vec{R}_{j})$, we can then perform a Fourier expansion on $\overline{D}_{\m\n\a\b}(\vec{R}_i - \vec{R}_j)$:
\begin{align}
    \overline{D}_{\m\n\a\b}(\vec{R}_i - \vec{R}_j) = \int \frac{\dd^2{\vec{p}}}{(2\pi)^2} \, e^{i \vec{p} \cdot (\vec{R}_i - \vec{R}_j)} D_{\m\n\a\b}(\vec{p}), \label{eqn:fourier_D}
\end{align}
Note that despite the two-center approximation, the momentum-space dynamical matrix $D_{\tilde{\vec{k}}\m\n\a\b}$ still depends on two different momenta from the opposite layers $\ell$ and $\ell'$.  
Using Eq.~\eqref{eqn:fourier_D}, as well as Eq.~\eqref{eq:sum} in Eq.~\eqref{eq:inter-doubleFT} yields 
\begin{widetext}
\begin{align}
\begin{aligned}
    D_{\widetilde{\vec{k}} \m\n\a\b}(\vec{k}^{(\ell)}, \vec{k}^{(\ell')}) & = \frac{1}{|\Gamma^*|} \int \frac{\dd^2{\vec{p}}}{(2\pi)^2} \; D_{\m\n\a\b}(\vec{p}) \sum_{ij} e^{-i (\vec{K}^{(\ell)} + \vec{p}) \cdot \vec{R}_{i}}  e^{i (\vec{K}^{(\ell')} + \vec{p}) \cdot \vec{R}_{j}} \\
    & = \frac{|\Gamma^*|}{(2\pi)^2} \sum_{\vec{G}^{(\ell)} \vec{G}^{(\ell')}} \int \dd^2{\vec{p}}\, D_{\m\n\a\b}(\vec{p})\, \d(\vec{K}^{(\ell)} + \vec{p} - \vec{G}^{(\ell)})\, \d(\vec{K}^{(\ell')} + \vec{p} - \vec{G}^{(\ell')}) \\
    & =\frac{1}{ |\Gamma|} \sum_{\vec{G}^{(\ell)} \vec{G}^{(\ell')}} D_{\m\n\a\b}(\widetilde{\vec{k}} + \vec{k}^{(\ell)} - \vec{G}^{(\ell)})\, \delta_{\vec{k}^{(\ell')} - \vec{k}^{(\ell)}, \vec{G}^{(\ell')} - \vec{G}^{(\ell)}}, \\
\end{aligned}
\end{align}
\end{widetext}
where $|\Gamma|$ is the area of the monolayer unit cell and we note that $|\Gamma| = (2\pi)^2 /|\Gamma^*|$. The $\delta$-function imposes a selection rule that constrains the allowed values of $\vecll{k}$ and $\vecl{k}{\ell'}$. 
For concreteness, we define each reciprocal lattice vector as \begin{align}
    \vec{G}^{(\ell)}_{mn} = m \vec{g}^{(\ell)}_1 + n \vec{g}_2^{(\ell)} ,
\end{align}
where $\vec{g}_1^{(\ell)}$ and $\vec{g}_2^{(\ell)}$ are a reciprocal lattice basis for layer $\ell$. Neglecting higher-order coupling, we assume that only terms with equal $m,n$ in $\vec{G}^{(\ell)}$ and $\vec{G}^{(\ell')}$ are nonzero, so the above simplifies to
\begin{widetext}
\begin{align}
    D_{\widetilde{\vec{k}}\m\n\a\b}(\vec{k}^{(\ell)},  \vec{k}^{(\ell')}) = \frac{1}{ |\Gamma|}\sum_{mn} & D_{\m\n\a\b}(\widetilde{\vec{k}} + \vec{k}^{(\ell)} - \vec{G}^{(\ell)}_{mn})\, \delta_{\vec{k}^{(\ell)} - \vec{k}^{(\ell')},  \widetilde{\vec{G}}_{mn}}.
\label{eq:inter-sel-rule}
\end{align}
\end{widetext}
The difference $\vec{G}^{(\ell)}_{mn} - \vec{G}^{(\ell')}_{mn}$ between corresponding reciprocal lattice vectors from each layer is a moir\'e reciprocal lattice vector $\widetilde{\vec{G}}_{mn}$.
Without loss of generality, taking $\ell=1$ and $\ell'=2$, Eq.~\eqref{eq:inter-sel-rule} imposes a constraint on the momenta given by
\begin{align} 
\vec{k}^{(1)} - \vec{k}^{(2)} = \vec{G}^{(2)}_{mn} - \vec{G}^{(1)}_{mn} = -\widetilde{\vec{G}}_{mn}. \label{eq:gmoire}
\end{align}
We henceforth drop the subscripts on $\widetilde{\vec{G}}$ for clarity. Equation~\eqref{eq:inter-sel-rule} is the fundamental equation governing the moir\'e dynamical matrix. The stacking-dependent part of the moir\'e phonon equation of motion becomes
\begin{widetext}
\begin{align}
\begin{aligned}
       \frac{1}{ |\Gamma|} \sum_{\n\b\ell'} \sum_{\widetilde{\vec{G}}} D_{\m\n\a\b}(\widetilde{\vec{k}} + \vec{k}^{(\ell)} - \vec{G}^{(\ell)})  \delta u_{\ell'\n\b}(\vec{K}^{(\ell')} -\widetilde{\vec{G}}) &= \omega^2  \delta u_{\ell\m\a}(\vec{K}^{(\ell)}), \label{eq:dynmat}
\end{aligned} 
\end{align}
\end{widetext}
where we note that $\widetilde{\vec{G}}$ and $\vec{G}^{(\ell)}$ are in one-to-one correspondence via Eq.~\eqref{eq:gmoire}. To simplify the above we choose a gauge $\vec{k}^{(\ell)} = \vec{G}^{(\ell')}$ (though other gauges will work, and we will choose a different one to justify the construction of the moir\'e dynamical matrix) with $\ell \neq \ell'$. Without loss of generality, we let $\ell = 1$:
\begin{align}
\begin{aligned}
    D_{\m\n\a\b}(\widetilde{\vec{k}} + \vec{k}^{(1)} - \vec{G}^{(1)}) & = D_{\m\n\a\b}(\vec{k} + \vec{G}^{(2)} - \vec{G}^{(1)}) \\
    & = D_{\m\n\a\b}(\widetilde{\vec{k}} - \widetilde{\vec{G}}) .
\end{aligned}
\end{align} 
Note that reciprocal space has sufficient symmetry that any choice of a pair of signs for $\vec{k}$ and $\widetilde{\vec{G}}$ produce the same term above. 

Combining the monolayer term (Eq.~\eqref{eqn:intra_eom}) and the twist-dependent term (Eq.~\eqref{eq:dynmat}), we obtain the moir\'e dynamical matrix. 
We verified numerically that the magnitude of the dynamical matrix decays rapidly with distance from the $\G$-point, so it suffices to expand up to the first shell of the moir\'e reciprocal lattice, which contains six vectors $\widetilde{\vec{G}}_1, \hdots, \widetilde{\vec{G}}_6$. For the off-diagonal terms, we ignore $\vec{\widetilde{k}}$-dependence. The interlayer term is explicitly given as follows:
\begin{align}
    D^{12} = \begin{bmatrix} 
 D_{\m\n\a\b}^{12}(\widetilde{\vec{G}}_{0}) & D_{\m\n\a\b}^{12}(\widetilde{\vec{G}}_{1}) & \cdots & D_{\m\n\a\b}^{12}(\widetilde{\vec{G}}_{6})\\
 D_{\m\n\a\b}^{12}(-\widetilde{\vec{G}}_{1}) & D_{\m\n\a\b}^{12}(\widetilde{\vec{G}}_{0}) & & \\ 
 \vdots & & \ddots & \\ 
 D_{\m\n\a\b}^{12}(-\widetilde{\vec{G}}_{6}) & & &  D_{\m\n\a\b}^{12}(\widetilde{\vec{G}}_{0}) 
 \end{bmatrix},
\end{align}
where $D^{ij}_{\m\n\a\b}(\widetilde{\vec{G}})$ is a submatrix of the dynamical matrix at $\widetilde{\vec{G}}$ with rows (columns) corresponding to degrees of freedom due to layer $i$ ($j$). The $(i, j)$ block thus represents phonon hoppings between reciprocal lattice sites $\vec{\widetilde{G}}_i$ and $\vec{\widetilde{G}}_j$. We have disregarded a small number of first-shell terms above that do not live along the first row, first column, or the diagonal, after having verified that their inclusion negligibly affects the numerical results. That is, we ignore hoppings that do not include $\vec{\widetilde{G}}_0$. Similarly for the intralayer term, 
\begin{align}
    D^i(\vec{\widetilde{k}}) = \begin{bmatrix} 
 D_P^{i}(\vec{\widetilde{k}}+\widetilde{\vec{G}}_0) & D_{\m\n\a\b}^{ii}(\widetilde{\vec{G}}_{1}) & \cdots & D_{\m\n\a\b}^{ii}(\widetilde{\vec{G}}_{6})\\
 D_{\m\n\a\b}^{ii}(-\widetilde{\vec{G}}_1) & D_P^{i}(\vec{\widetilde{k}}+\widetilde{\vec{G}}_1) & & \\ 
 \vdots & & \ddots & \\ 
 D_{\m\n\a\b}^{ii}(-\widetilde{\vec{G}}_{6}) & & & D_P^{i}(\vec{\widetilde{k}}+\widetilde{\vec{G}}_6)
 \end{bmatrix},
\end{align}
where the diagonal terms are the dynamical matrix of the pristine monolayer $D^i_P(\vec{\widetilde{k}} + \vec{\widetilde{G}})$ and the off-diagonal terms being the twist-dependent intralayer term constructed from Eq.~\eqref{eq:dynmat} when $\ell=\ell'$.

\subsection{\label{sec:csc:sub:cfg} Dynamical Matrix in Configuration Space}
We are particularly interested in structures with a small twist angle, which means the moir\'e cell contains a large number of primitive cells and consequently many atoms. 
In order to make the model computationally feasible, we express $D_{\m\n\a\b}$ in terms of a collection of local atomic environments, known as configuration space \cite{carr2018relaxation}. 
Within the continuum approximation we replace the primitive cell lattice vectors $\mathbf{R}$ with a continuous variable $\vec{r}$ defined throughout the moir\'e cell, and allow quantities of interest to be defined at all $\vec{r}$. At each location $\vec{r}$ the local environment can be approximated by an untwisted bilayer with a horizontal shift of one layer relative to the other. We specify the interlayer shift by $\vec{b}(\vec{r})$, and the space formed by all $\vec{b}$ is configuration space. It is in one-to-one correspondence with real space by a linear transformation
\begin{align}
    \vec{b}(\vec{r}) = (\mathbb{1} - A_2 A_1^{-1}) \vec{r},\label{eqn:mapping}
\end{align}
where $A_i$ is a matrix whose columns are the primitive lattice vectors of layer $i$ and $\mathbb{1}$ is the $2\times2$ identity matrix. 
This correspondence also gives a useful relationship between the the reciprocal lattice vectors of the moir\'e cell, $\vec{\widetilde{G}}$, and those of the primitive unit cell, $\vec{G}$: 
\begin{align}
    \vec{\widetilde{G}} \cdot \vec{r} = \vec{G} \cdot \vec{b}(\vec{r}).
\end{align} 
Using these concepts, we can express the momentum-space dynamical matrix as follows:
\begin{align}
    \begin{aligned}
        D^{\text{rig}}_{\m\n\a\b}(\widetilde{\vec{k}} + \widetilde{\vec{G}}) = & \frac{1}{\sqrt{|\G^*|}}
        \int d^2\vec{r}\,  \overline{D}_{\m\n\a\b}(\vec{r})\, e^{i(\tilde{\vec{k}}+\tilde{\vec{G}})\cdot\vec{r}}\\
        =& \frac{1}{\sqrt{|\G^*|}} \int d^2\vec{b} \, \overline{D}_{\m\n\a\b}(\vec{b})\, e^{i(\vec{k}+\vec{G})\cdot\vec{b}} \\
        \approx& \frac{1}{\sqrt{|\G^*|}} \frac{|\G|}{N^2} \sum_{\vec{b} \in \mathcal{S}(N)} \overline{D}_{\m\n\a\b}^{(\vec{b})}\, e^{i (\vec{k} + \vec{G}) \cdot \vec{b}},
    \end{aligned}
\end{align}
The factor of $|\G|$, arising from discretization, cancels out the normalization from Eq.~\eqref{eq:inter-sel-rule}.
Configuration space allows us to replace $\overline{D}(\vec{r})$, which we cannot easily determine, with $\overline{D}(\vec{b})$, which we can readily compute with DFT for a primitive bilayer cell with any specified shift $\vec{b}$.
In the small $\theta$ limit the local environment varies slowly and smoothly with $\vec{b}$, so we can approximate the integral over $\vec{b}$ with the sum
over a $N \times N$ uniform mesh $\mathcal{S}(N)$ of $\vec{b}$ values.
Since $\mathcal{S}(N)$ is uniform, $D^{\text{rig}}_{\m\n\a\b}(\widetilde{\vec{k}} + \widetilde{\vec{G}})$ describes a moir\'e bilayer with a uniform distribution of local configurations. In other words, it represents a rigid moir\'e structure with no twist-induced continuum relaxation, which is a good approximation for $\th$ sufficiently large. 
We describe the procedure for including relaxation effects in the next subsection.

\subsection{\label{sec:csc:sub:reldm}Continuum Relaxation}
We emphasize that in all structures considered here atomic relaxation to the lowest-energy (equilibrium) configuration is included. However, its effect varies as a function of twist angle: for
moderate to small twist angles, twist-induced atomic relaxation
is much more important than for large twist angles. The range 
of angles over which the atomic relaxation is important
varies by material, but relaxation typically begins to have a significant impact
for twist angles smaller than
about $2^\circ$. In real space, atomic relaxation results in enlarged areas of the lowest energy stacking configurations (AB and BA in graphene) at the expense of high energy stackings (AA), with domain walls forming in between.
In configuration space relaxation can be viewed as a 
higher density of configurations in the lowest-energy stackings~\cite{carr2018relaxation,triGr}. 
This can be accounted for by shifting each $\vec{b}$ in the uniform mesh $\mathcal{S}(N)$ by an angle-dependent relaxation displacement vector $\vec{u}_\th(\vec{b})$, yielding 
a non-uniform mesh $\vec{B} = \vec{b} + \vec{u}_\th(\vec{b})$ denoted $\mathcal{S}_\th(N)$. 
We use linear elasticity theory to describe the intralayer energy, combined with the generalized stacking fault energy (GSFE) functional for the interlayer energy. (The GSFE parameters are given in Appendix \ref{sec:GSFE}.) By minimizing the total energy we can compute $\vec{u}_\th(\vec{b})$ in configuration space.
We refer to \citet{carr2018relaxation} and \citet{triGr} for a detailed overview of continuum relaxation.
The dynamical matrix can then be updated by Fourier interpolation, using an inverse transform
\begin{align}
    \overline{D}_{\m\n\a\b}^{(\vec{B})}(\vec{\widetilde{k}}) = \frac{1}{\sqrt{|\Gamma^*|}} \sum_{\vec{\widetilde{G}}} D^{\text{rig}}_{\m\n\a\b}(\widetilde{\vec{k}} + \widetilde{\vec{G}}) e^{-i(\vec{k} + \vec{G}) \cdot \vec{B}},
\end{align}
followed by a second forward transform, which arrives at the following relaxed moir\'e dynamical matrix:
\begin{align}
    D_{\m\n\a\b}(\widetilde{\vec{k}} + \widetilde{\vec{G}}) = \frac{1}{\sqrt{|\G^*|}} \frac{|\G|}{N^2} \sum_{\vec{b} \in \mathcal{S}(N) } \overline{D}_{\m\n\a\b}^{(\vec{B})}(\widetilde{\vec{k}}) e^{i (\vec{k} + \vec{G}) \cdot \vec{b}}.
\end{align}

To obtain the real space phonon vibration pattern, we perform an inverse Fourier transform by summing over components of the eigenvectors as follows,  
\begin{align}
    \d u_{n\ell\m\a\widetilde{\vec{k}}}(\vec{r}) = \sum_{\vec{\widetilde{G}}} \d u_{n\ell\m\a\widetilde{\vec{G}}}(\widetilde{\vec{k}}) e^{i (\vec{\widetilde{G}} + \vec{\widetilde{k}}) \cdot \vec{r}},
    \label{eq:inverse_FT}
\end{align}
where $n$ is the band index and $\d u_{n\ell\m\a\widetilde{\vec{G}}}$ is the eigenvector component that corresponds to $\widetilde{\vec{G}}$.

\subsection{Acoustic Sum Rules} \label{app:sum_rule}
Even though the total forces are fixed to be zero in the force field calculations of individual stackings, the combination of all configurations in the moir\'e dynamical matrix construction modifies the total force of the combined system.
Therefore, an acoustic sum of the moir\'e dynamical matrix is required to ensure the total force stays zero in addition to the sum rules on the unrotated bilayers and monolayers.
We use a correction matrix (chosen without loss of generality to be diagonal in the latter pair of indices) \begin{align}
    C_{\m\n\a\a} = \sum_{\b} \sqrt{\frac{M_\b}{M_\a}} D_{\text{moir\'e},\m\n\b\a}^*(\G) ,
\end{align}
where $D_{\text{moir\'e},\m\n\b\a}^*(\G)$ is the moir\'e dynamical matrix by replacing the pristine monolayer blocks $D_P(\vec{k} + \vec{\widetilde{G}})$ with $D_P(\G)$, as we enforce translational invariance only at the $\G$ point and enforce continuity everywhere else. We then adjust $D_{\text{moir\'e}}(\vec{\widetilde{k}}) \leftarrow D_{\text{moir\'e}}(\vec{\widetilde{k}}) - C$, observing numerically that the above approximation satisfies adequately the translational invariance constraint. The sum rule is adapted from \citet{quan2021phonon}.

\section{\label{sec:comparison}Model Validation}
To validate our CSC model, we compare results obtained from it with two first-principles models, namely direct DFT on the moir\'e supercell for a small system where such calculations are feasible, and with molecular dynamics (MD) simulations that employ an interatomic potential~\cite{Fabrizio_PRX2019, Fabrizio_PRB2018}.  We also compare our CSC results to another empirical continuum (EC) model adapted from \citet{koshino2019moire}.

\subsection{\label{sec:model_comparison}Comparison with Molecular Dynamics and DFT}
Direct DFT phonon calculations on the moir\'e cell are the most accurate but are restricted by the size of the cell. MD simulations allow for larger supercell sizes, but have inherent limitations on accuracy due to the choice of interatomic potential. Both direct DFT and MD are computationally expensive and require a commensurate supercell. Since the CSC model uses DFT inputs in a well-defined approximation, it should reflect the direct DFT and MD calculations reasonably well and does not suffer constraints based on moir\'e supercell size or commensurability.
We choose a twist angle, $\th = 7.34^\circ$, that gives a commensurate pattern in bilayer graphene for these comparisons.

\begin{figure*}[ht!]
    \centering
    \includegraphics[width=\linewidth]{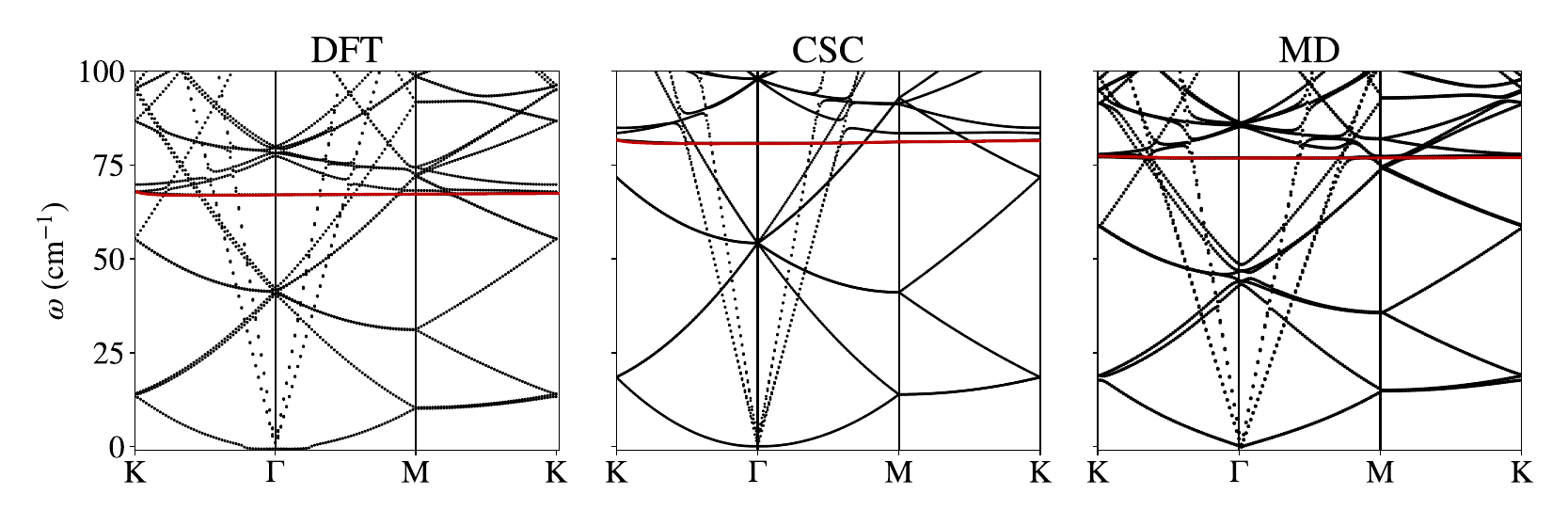}
    \caption{Comparison of phonon band at $\theta = 7.34^\circ$ between: DFT (left panel), CSC (middle panel), and MD (right panel) calculations for graphene. The LB mode is highlighted in red.}
    \label{fig:3-Comparison}
\end{figure*}

Figure \ref{fig:3-Comparison} compares the phonon bands of the CSC model, MD simulations, and direct DFT phonon calculations on the moir\'e supercell at $\th = 7.34^\circ$. The layer breathing (LB) mode, which represents layer 1 (2) moving in the $-z$ ($+z$) direction and a much weaker in-plane motion, is highlighted in red in each case.
The results from the three models are qualitatively
quite similar, validating the CSC construction. The quantitative differences, such as the exact frequency of the LB mode, are to be expected due to the specific choices made for each model, such as the exchange correlation functional for DFT and the interatomic potential for MD.

\begin{figure*}[ht!]
    \centering
    \includegraphics[width=5in]{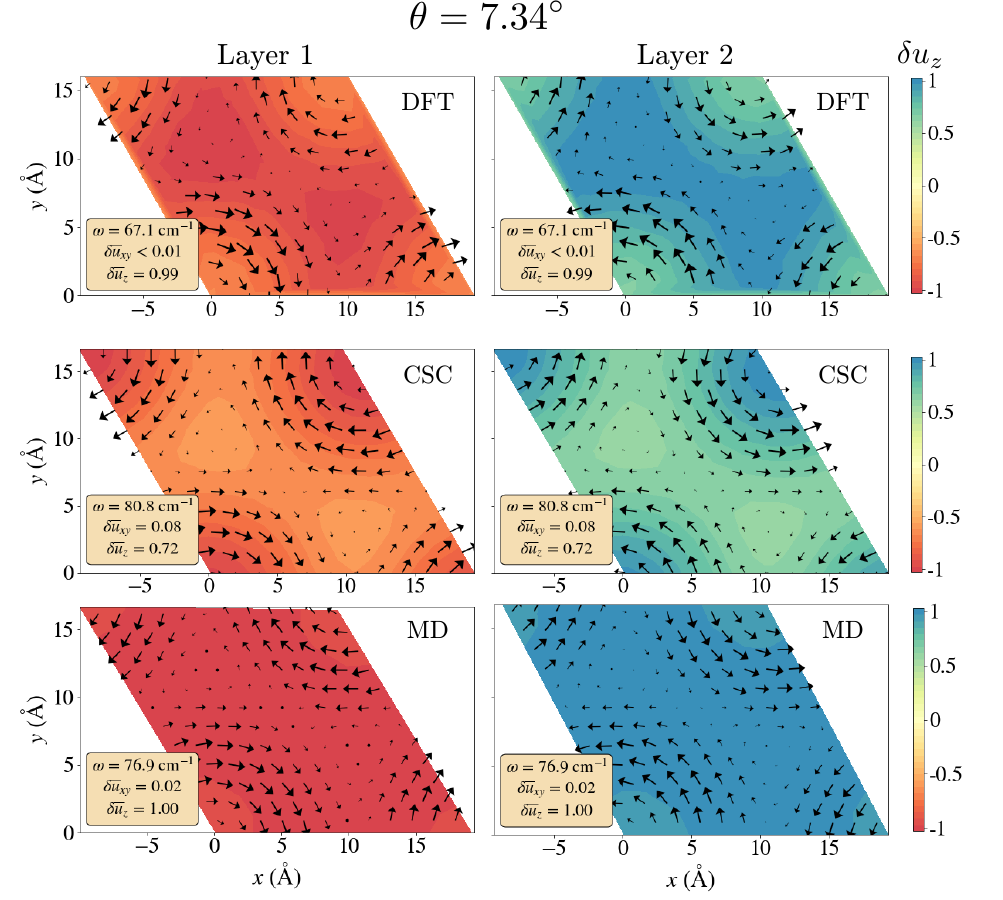}
    \caption{Comparison of the LB atomic displacements between direct DFT, CSC, and MD. Each model has normalized its displacements $\d \vec{u}$ separately, in arbitrary units. The insets show the frequency and average displacement magnitude in-plane ($\d \overline{u}_{xy}$) and out-of-plane ($\d \overline{u}_z$).}
    \label{fig:2-MDRS}
\end{figure*}

In addition to the phonon dispersion relations, Fig.~\ref{fig:2-MDRS} shows a comparison of the atomic displacements of the LB mode in graphene, corresponding to the red band in Fig.~\ref{fig:3-Comparison} at the $\Gamma$-point. We focus on a qualitative comparison of the geometrical structure. All models exhibit a swirling in-plane motion around the AA regions at the corner of the moir\'e supercell. The primary difference between the models is the variation in the out-of-plane motion. In the DFT calculation the maximum vertical displacement occurs in the AB regions, whereas for the CSC model the maximum is in the AA regions. While the MD calculation shows some variation in the vertical displacement between AA and AB spots and the maximum vertical displacement also occurs in the AB regions as in DFT, it is mostly uniform throughout the moir\'e cell.
The reason for the quantitative difference between the CSC and the first-principles models is likely due to the breakdown of the continuum approximation at a large angle.
When the twist angle is large, as is the case for $\theta=7.34^\circ$ in graphene, the local stacking order varies drastically from one real space position to a neighboring one. 
However, the CSC still assumes a smooth variation of the force fields in real space, which causes the real space LB mode to be similar for all twist angles and thus it fails to capture the change in the large angle.
Despite this slight disagreement, both the LB frequency and the band structure from the CSC show an excellent agreement with first-principles calculations (Fig.~\ref{fig:2-MDRS}).
There agreement in the LB mode out-of-plane displacement field is better at smaller twist angles (not shown in Fig.~\ref{fig:2-MDRS}), that is, all three models exhibit maxima in the displacement field around the AA regions.

\subsection{\label{sec:details}Comparison with Empirical Continuum Model}
\begin{figure*}[ht!]
    \centering
    \includegraphics[width=\linewidth]{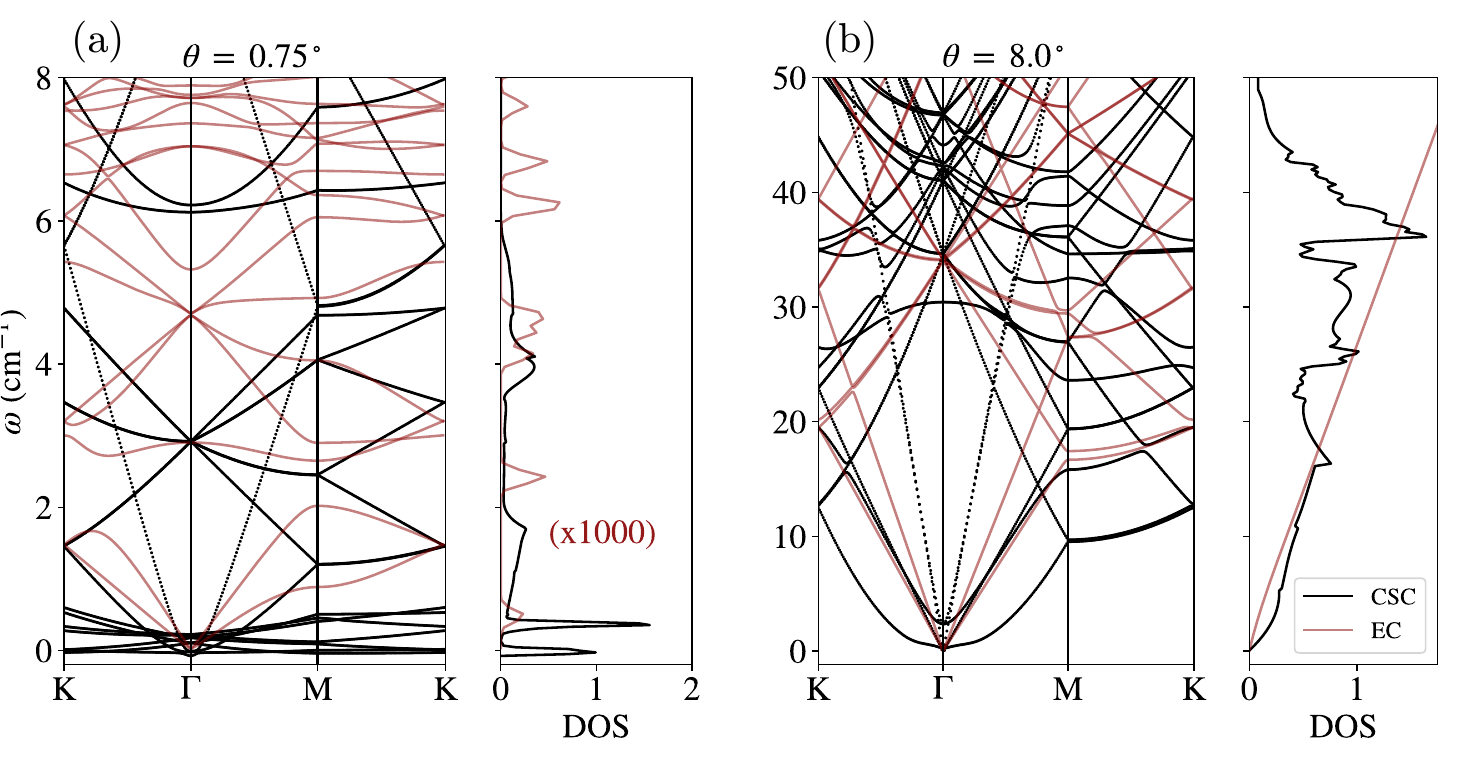}
    \caption{Comparison of the phonon bands and DOS for bilayer MoS$_2$ at (a) $0.75^\circ$ and (b) $8.0^\circ$ as calculated using the EC model (red lines) and the CSC model (black lines). For $\theta = 0.75^\circ$, the EC DOS has been scaled by a factor of 1000 for visualization.}
    \label{fig:4-Koshino}
\end{figure*}

The EC model, based on empirical interlayer interactions, is the most efficient of the models discussed in this section~\cite{koshino2019moire}. In Fig. \ref{fig:4-Koshino} we compare the phonon dispersion relations and density of states for MoS$_2$, obtained with the EC model and with our CSC model. 
To obtain the in-plane phonon modes, we perturb around the relaxed equilibrium positions by adding a kinetic energy contribution to the interlayer and intralayer terms of the dynamical matrix. 
We then solve the equation of motion to obtain the phonon modes.
We use the GSFE coefficients and elastic constants provided in Tables \ref{tab:GSFE} and \ref{tab:elastic} for the CSC model.
A detailed derivation is given in Appendix~\ref{sec:koshino}.

 The EC model has a free empirical parameter that we determine by scaling the bands such that the lowest-frequency in-plane translational modes at K coincide in the two models; the scaling factor is 2.3 for $\theta = 0.75^\circ$ and 22.5 for $8.0^\circ$. 
 
There is little agreement between the two models for small $\th$, as shown in Fig. \ref{fig:4-Koshino}(a). The EC model predicts flatter bands and exhibits a gap between 2 and 3 cm$^{-1}$, whereas no such features occur in the CSC model. Moreover, the slopes of the in-plane acoustic modes cannot be matched well. Intrinsically, the EC model also does not include any out-of-plane motion, such as the $z$-translational mode, nor does it include interlayer modes such as the shearing (S) mode. The LB mode is absent in the EC model for these two reasons. Our continuum model does not suffer from such limitations.

In the large twist-angle limit, Fig. \ref{fig:4-Koshino}(b), the results from the two models agree for low frequencies near $\G$; they both exhibit linear DOS relations and the position of the first folded band cluster is near 35 cm$^{-1}$.

\begin{figure*}[ht!]
    \centering
    \includegraphics[width=\linewidth]{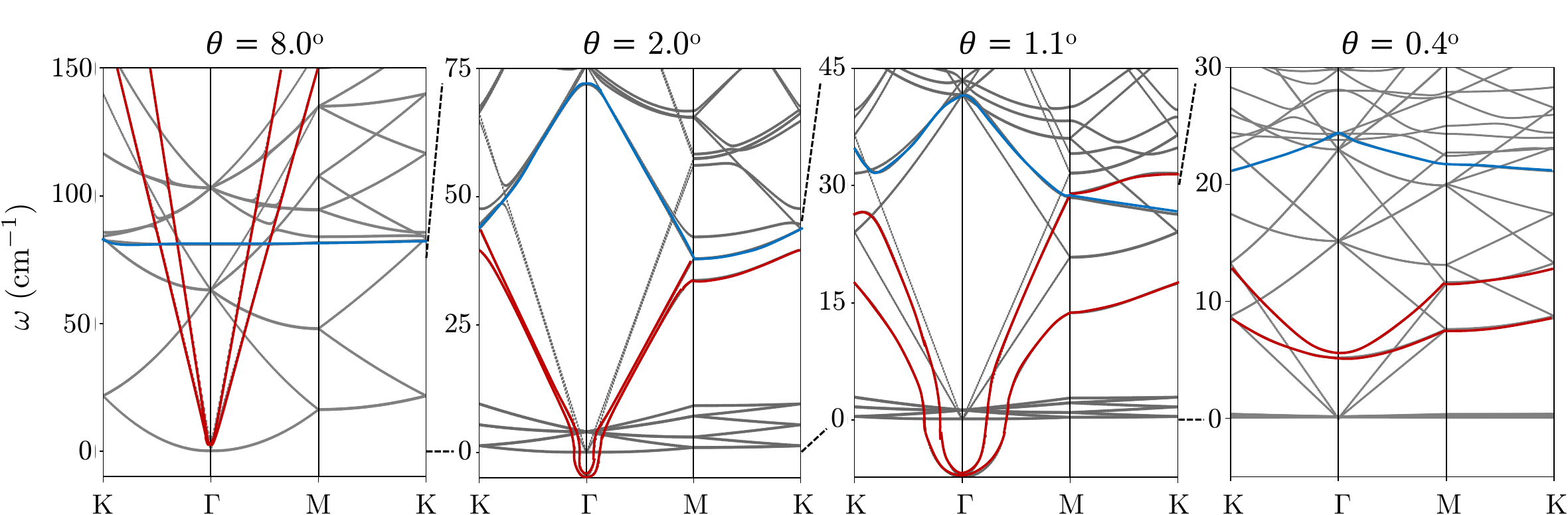}
    \caption{Phonon bands of graphene as obtained from our CSC model at four representative angles of (a) $8^\circ$, (b) $2^\circ$, (c) $1.1^\circ$, and (d) $0.4^\circ$. The two shearing modes are highlighted in red, and the first (lowest energy) LB mode in blue.}
    \label{fig:4-bands}
\end{figure*}
\section{\label{sec:results}Moir\'e Phonon Properties}
Using our CSC model, we calculate the phonon properties of three different moir\'e bilayers: graphene, MoS$_2$, and a MoSe$_2$-WSe$_2$ heterostructure.
Near the $\Gamma$-point the lowest energy modes include two shearing modes (S) and one layer breathing (LB) mode. These represent in-plane and out-of-plane motions, respectively, with each layer moving in opposite direction.


\subsection{Low-energy mode frequencies}\label{sec:res:sub:band}

The phonon dispersion relations are qualitatively similar for twisted bilayer graphene and TMDCs. Here, we focus on graphene.
Figure~\ref{fig:4-bands} shows the graphene phonon bands at different twist angles $\th$. 
As $\th$ decreases, the frequency where folded clusters of phonon bands appear also decreases, due to the shrinking of the moir\'e Brillouin zone area. 
Atomic relaxation due to interlayer interactions becomes more significant as the twist angle decrease, leading to three qualitatively different regimes characterized by the value of $\th$:
a ``decoupled'' regime (large $\th$), a ``soft'' regime (moderate $\th$) and a ``relaxed'' regime (small $\th$). 
The exact values of $\theta$ separating these regimes are specific to the material. 
For graphene they are approximately $4.2^\circ$ and $0.5^\circ$, respectively; for MoS$_2$ they are $7.6^\circ$ and $1^\circ$; and for MoSe$_2$-WSe$_2$ they are $5.5^\circ$ and $1.0^\circ$. 
In the decoupled regime (Fig.~\ref{fig:4-bands}(a)) the relatively large twist angle and small moir\'e scale means there is little atomic relaxation and the net interlayer coupling is very weak. Here, the S mode is slightly positive at $\G$ with twofold degeneracy (when the distinction between the two S modes becomes important, we shall refer to the lower-energy curve as S$_1$ and the higher-energy curve as S$_2$), while the LB mode is essentially dispersionless. The zero-frequency modes at $\G$ are the three translational modes. In the soft regime (Fig.~\ref{fig:4-bands}(b)), where atomic relaxation starts to become more noticeable, the S mode begins to exhibit imaginary frequencies at the $\Gamma$ point (plotted as negative values for continuity in the phonon bands). In the relaxed regime (Fig.~\ref{fig:4-bands}(d)) the low-energy stacking regions grow in size at the expense of the high-energy stacking regions. Thus, most of the atoms are in local energy minima and are therefore resistant to shearing. As a result, the shearing 
mode frequency becomes positive again.

Such ultrasoft S modes have been discussed in previous work~\cite{maity2020phonons}. In the limit of small angles (large moir\'e cells) that are truly incommensurate, they are zero-modes at $\Gamma$ that correspond to the invariance of the moir\'e pattern to a horizontal shift of one layer relative to the other~\cite{bistritzer2011moire,ochoa2019moire}. Given the complicated structure of the moir\'e dynamical matrix, the delicate numerical cancellations necessary to reproduce a (near) zero-mode are unlikely to occur in practice. For comparison, in standard phonon calculations the zero-modes (acoustic phonons, particularly the translational mode with $z$ invariance in 2D materials) are corrected by imposing an acoustic sum rule that explicitly enforces translational invariance. As discussed previously, we impose such a rule in the CSC model as well, which is why there are no negative-frequency acoustic phonon bands. In the same spirit, to deal with the ultrasoft shearing modes one could impose a ``shearing sum rule" to enforce this additional symmetry of moir\'e structures. Such a rule, however, is more subtle and involved than the acoustic sum rule, and we leave its imposition to future work. The imaginary frequencies are not a simple consequence of numerical sampling error or relaxation effects, as we demonstrate in Appendix~\ref{app:soft-regime}.

In the soft regime, the LB mode becomes dispersive, as seen in Fig.~\ref{fig:4-bands}(b)-(c), but progressively flattens out again as $\th$ approaches the relaxed regime, Fig.~\ref{fig:4-bands}(d). At $1.1^\circ$, the LB mode has a crossing with a cluster of folded bands whose frequency decreases 
as the Brillouin zone shrinks (\ref{fig:4-bands}(c)), and similarly for $\theta=0.4^\circ$ (\ref{fig:4-bands}(d)).

\begin{figure*}[ht!]
    \centering
    \includegraphics[width=\linewidth]{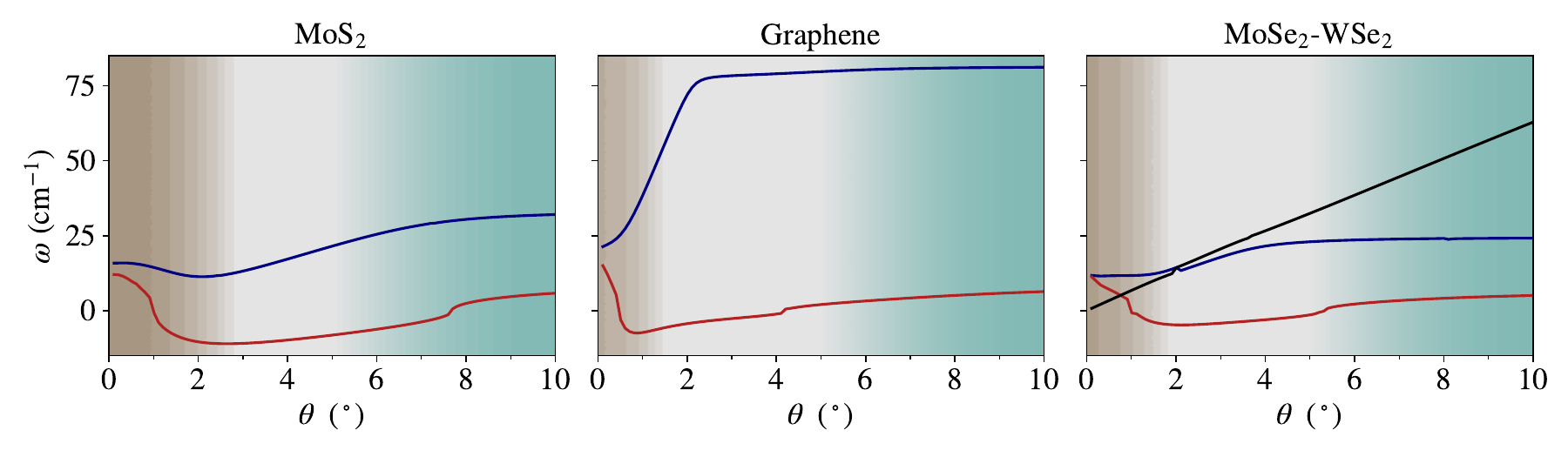}
    \caption{Frequency of the shearing (red line) and layer breathing (blue line) modes at $\G$ as a function of twist angle $\th$ for various bilayers. The MoSe$_2$-WSe$_2$ heterostructure has an additional LB-like mode that breaks symmetry, labeled SBLB (black line). The background color shades illustrate the transition from the decoupled region (large $\th$, S frequencies positive or nearly 0) in blue shade, to the soft region (from initial dip to negative S frequencies to angle of maximally negative frequency) in white shade, to the relaxed region (upwards curve to positive S frequencies) in brown shade.}
    \label{fig:2-modes}
\end{figure*}

We may examine the frequency dependence on twist angle more thoroughly by focusing on the $\G$ point. Figure~\ref{fig:2-modes} compares the S and LB frequencies for graphene, $\mathrm{MoS_2}$, and MoSe$_2$-WSe$_2$. The decoupled, soft, and relaxed regimes are shaded blue, white, and brown, respectively. The soft regime begins when the S mode takes negative frequency values, and ends where the slope of $\w$ becomes negative at small $\th$. Aside from the location and width of the soft regime, however, the frequency dependence on $\theta$ of the S mode is qualitatively similar for all three materials. For MoS$_2$ and MoSe$_2$-WSe$_2$, the LB varies smoothly and slowly with $\th$. By contrast, the LB mode in graphene has a comparatively higher frequency in the decoupled regime, which drops sharply near the end of the soft regime. Since the MoSe$_2$-WSe$_2$ heterostructure breaks the layer symmetry, it admits an additional LB-like mode that does not respect the symmetry under layer inversion ($\vec{r}\rightarrow -\vec{r}$). We refer to this mode as the symmetry-breaking LB (SBLB) mode, and note that it is the only mode that varies linearly in frequency with respect to $\theta$.

\subsection{Real Space Atomic Displacements at the $\G$ Point}\label{sec:res:subsec:thetaspc}
By performing the inverse Fourier transform of eigenmodes in reciprocal space (Eq.~\eqref{eq:inverse_FT}), we obtain the atomic displacement field of each phonon mode in real space.
The eigenmodes in reciprocal space are unit vectors, but application of the Fourier transform gives the magnitude of the displacement vectors a physical interpretation. 
Specifically, the overall magnitudes across various twist angles serve as an indicator for the magnitude of each eigenmode component $\d \vec{u}(\widetilde{\vec{G}})$ in reciprocal space. For if most of the magnitude in reciprocal space is concentrated in the $\widetilde{\vec{G}}_0$ component, the real space magnitude will be large; but if most of the magnitude is concentrated in the first shell $\widetilde{\vec{G}}_1, \hdots, \widetilde{\vec{G}}_6$, the phases in the inverse Fourier transform reduce the real space magnitude to a much smaller quantity. In the absence of the moir\'e interlayer couplings, the magnitude in Fourier space should be concentrated in the $\widetilde{\vec{G}}_0$ component, but at smaller values of $\th$, emergence of periodicity of the moir\'e length scale imply that the magnitude becomes instead concentrated in higher shells. Thus, we should expect the magnitude in real space to fall substantially near the angle at which the magnitudes of the $\widetilde{\vec{G}}_0$ component and the $\widetilde{\vec{G}}_1, \hdots, \widetilde{\vec{G}}_6$ components intersect. 

\begin{figure*}[ht!]
    \centering
    \includegraphics[width=\linewidth]{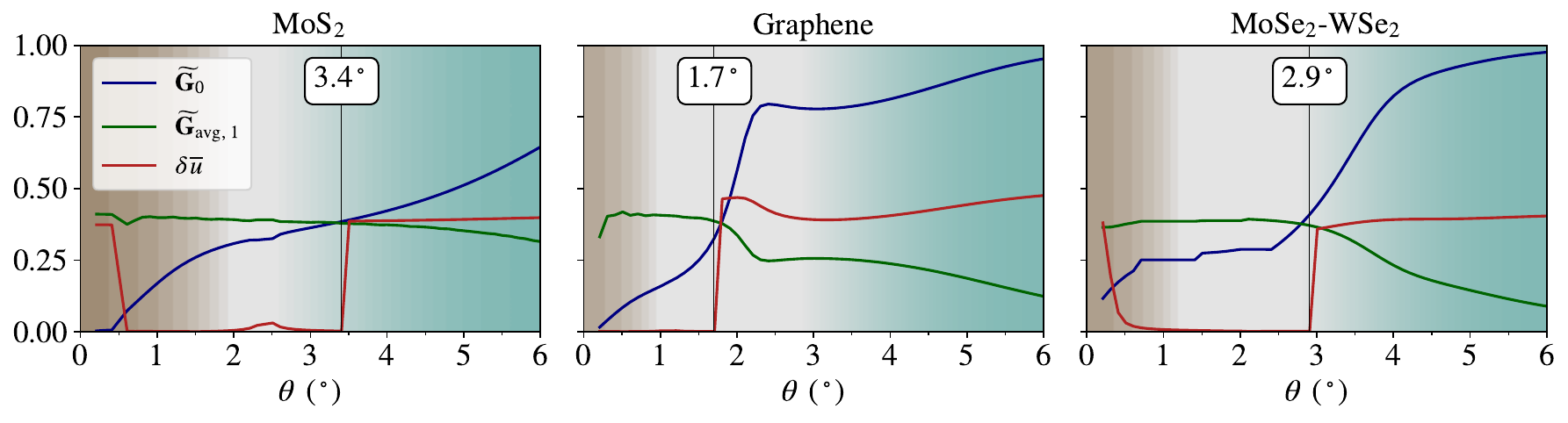}
    \caption{Average atomic displacement magnitudes in real space ($\d \overline{u}$, red), of the $\widetilde{\vec{G}}_0$ component in reciprocal space (blue), and of the average of the first-shell components $\widetilde{\vec{G}}_1, \hdots, \widetilde{\vec{G}}_6$ of the LB mode at $\G$ for MoS$_2$, graphene, and MoSe$_2$-WSe$_2$. The shading of various regimes has the same meaning as in Fig.~\ref{fig:2-modes}. The vertical black lines indicate the largest value of the twist angle at which $\d\overline{u} \to 0$.}
    \label{fig:3-mags}
\end{figure*}

Figure~\ref{fig:3-mags} demonstrates these features in the LB mode by approximating the average real space magnitude through a $13 \times 13$ mesh discretization of the moir\'e supercell. For brevity, we have averaged the magnitudes of the six first-shell reciprocal lattice vectors. 
Note that the magnitude of the $\widetilde{\vec{G}}_0$ component can be measured via Raman spectroscopy~\cite{quan2021phonon}, and thus the real space magnitude across $\th$ serves to indicate at which twist angle the mode will be difficult to detect experimentally.

The magnitudes of $\d\vec{u}(\vec{r})$ also allow for comparison of the relative strength between in-plane ($\d u_{xy}$) and out-of-plane ($\d u_z$) motion; for example, the S modes satisfy $\d u_z \ll \d u_{xy}$. This is seen explicitly by visualization of the S and LB atomic displacement fields, from which we may also deduce the geometric structure of the atomic displacements as a function of $\th$. In all figures showing atomic displacement field results from the CSC model, we normalize the magnitudes that correspond to each atomic degree of freedom by the atomic mass $\sqrt{M_\a}$ for each layer, and subsequently average the vectors over the atomic degrees of freedom. 
We verified numerically that for all modes analyzed in the present work, the vectors for each atomic degree of freedom are identical up to mass normalization, which justifies the averaging.
We then renormalize by multiplying the atom-averaged field by $\sqrt{\sum_\a M_\a}$, where $\alpha$ is the atomic index. 

\begin{figure*}[ht!]
    \centering
    \includegraphics[width=5in]{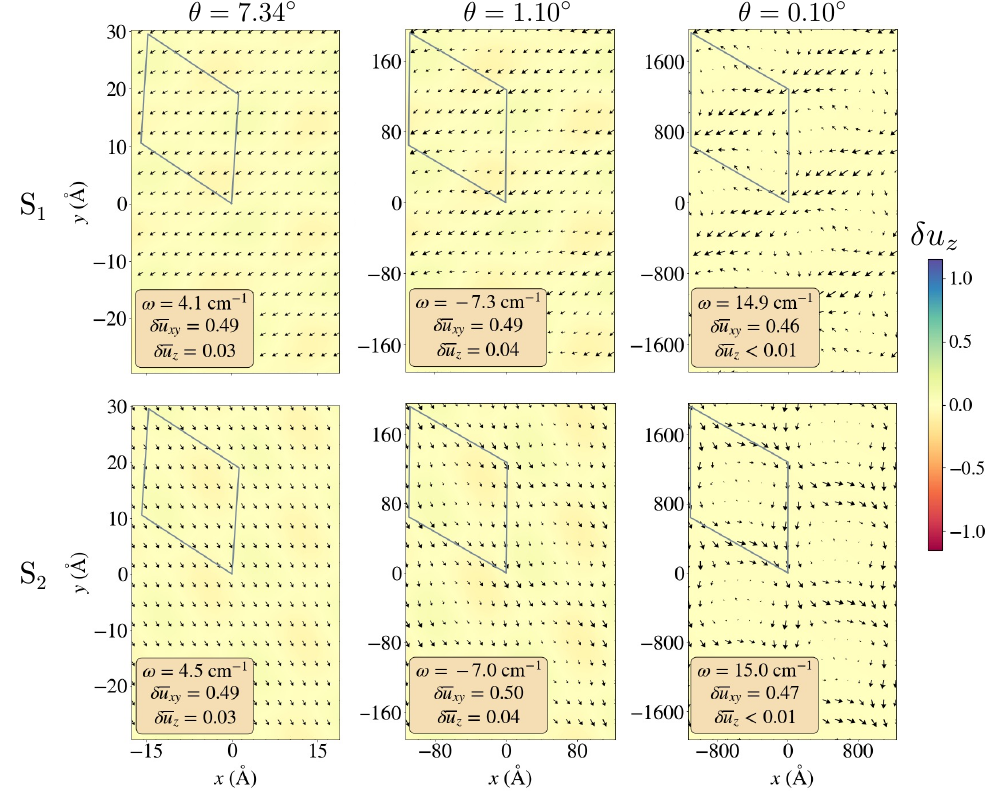}
    \caption{Real space atomic displacements (in arbitrary units) for the twofold degenerate shearing modes, S$_1$ and S$_2$, of the first layer of graphene at various twist angles. In-plane ($x,y$) displacements are proportional to the arrows and vertical ($z$) displacements are shown in color. The arrows are rescaled for visualization purposes, with the average values of in-plane ($\d \overline{u}_{xy}$) and out-of-plane ($\d \overline{u}_z$) magnitudes, rounded to 2 decimal places, shown in the lower insets. The displacements of the layer 2 are exactly opposite to those of layer 1. The grey parallelogram outlines a single moir\'e cell.}
    \label{fig:7-shear}
\end{figure*}

Figure~\ref{fig:7-shear} illustrates the evolution of the atomic displacement field of the two S modes, S$_1$ and S$_2$, as a function of $\theta$. 
In all cases, the S mode is dominated by in-plane displacements and is opposite in the two layers. 
In the decoupled regime ($7.34^\circ$) the displacement is uniform. 
As $\theta$ decreases, translational symmetry on the graphene unit cell scale is broken and we observe structure at the moir\'e scale. 

\begin{figure*}[ht!]
    \centering
    \includegraphics[scale=0.5]{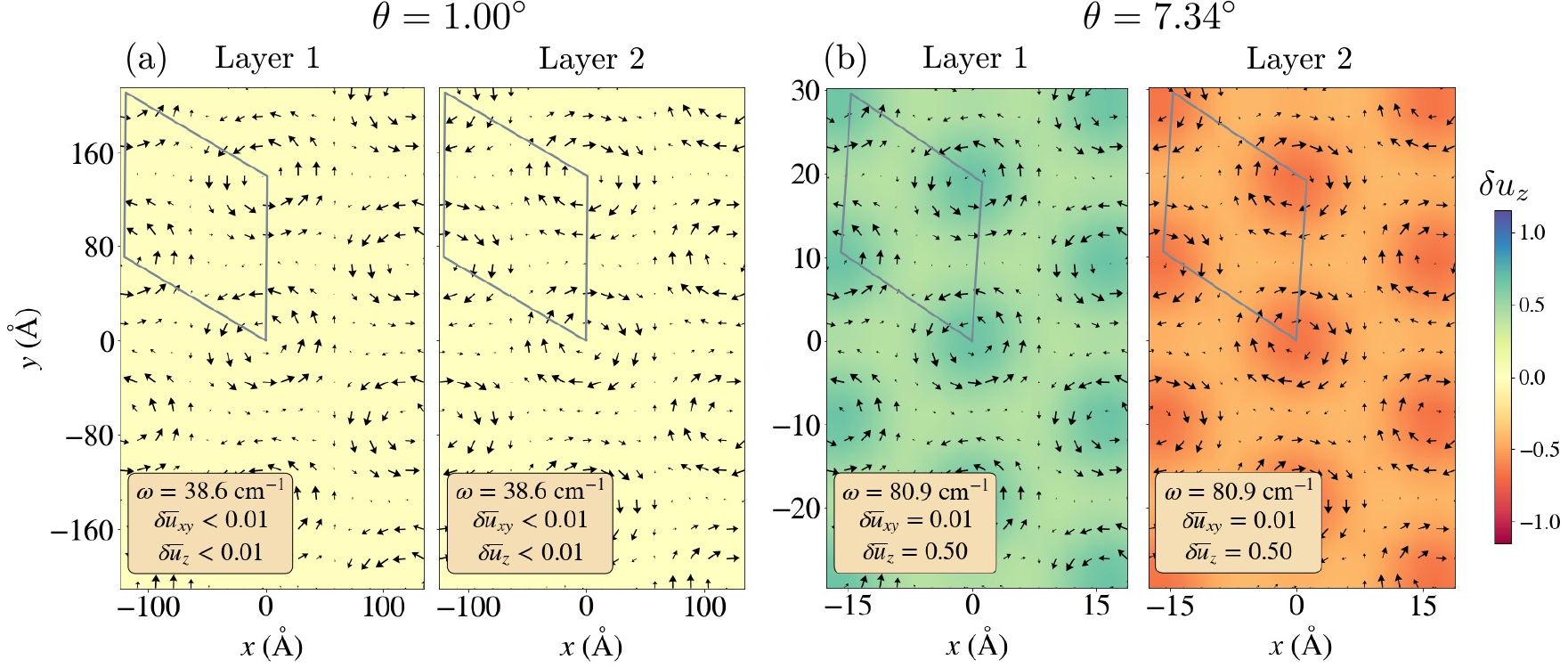}
    \caption{LB atomic displacements of graphene at twist angles (a) $1.00^\circ$ and (b) $7.34^\circ$. Color scale and insets are the same as in Fig.~\ref{fig:7-shear}. For each angle, the left (right) panel shows layer 1 (2).}
    \label{fig:5-LB}
\end{figure*}

In contrast to the S modes, the geometrical structure of the LB mode remains invariant as $\th$ changes. Figure~\ref{fig:5-LB} shows the LB modes at two different angles in which the two layers have equal and opposite displacements along the $z$-direction, similar to the LB mode in AB bilayer graphene. However, while the LB mode in an untwisted bilayer system corresponds to a negligible in-plane motion, the moir\'e LB modes correspond to swirling motion around the corners of the moir\'e supercell in opposing directions between layers, which resembles the relaxation pattern in twisted bilayer graphene~\cite{carr2018relaxation}.
This is because the $\Gamma$ point LB mode changes the interlayer separation on the moir\'e scale, which modifies the local interlayer energy.
To compensate for such an energy cost, the mode adopts an in-plane rotation that changes the total AAarea. Consistent with Fig.~\ref{fig:3-mags}, the relative magnitude of the LB mode displacement field varies substantially with respect to twist angle, as shown in the insets of Fig. \ref{fig:5-LB}.

\begin{figure}[ht!]
    \centering
    \includegraphics[width=3.25in]{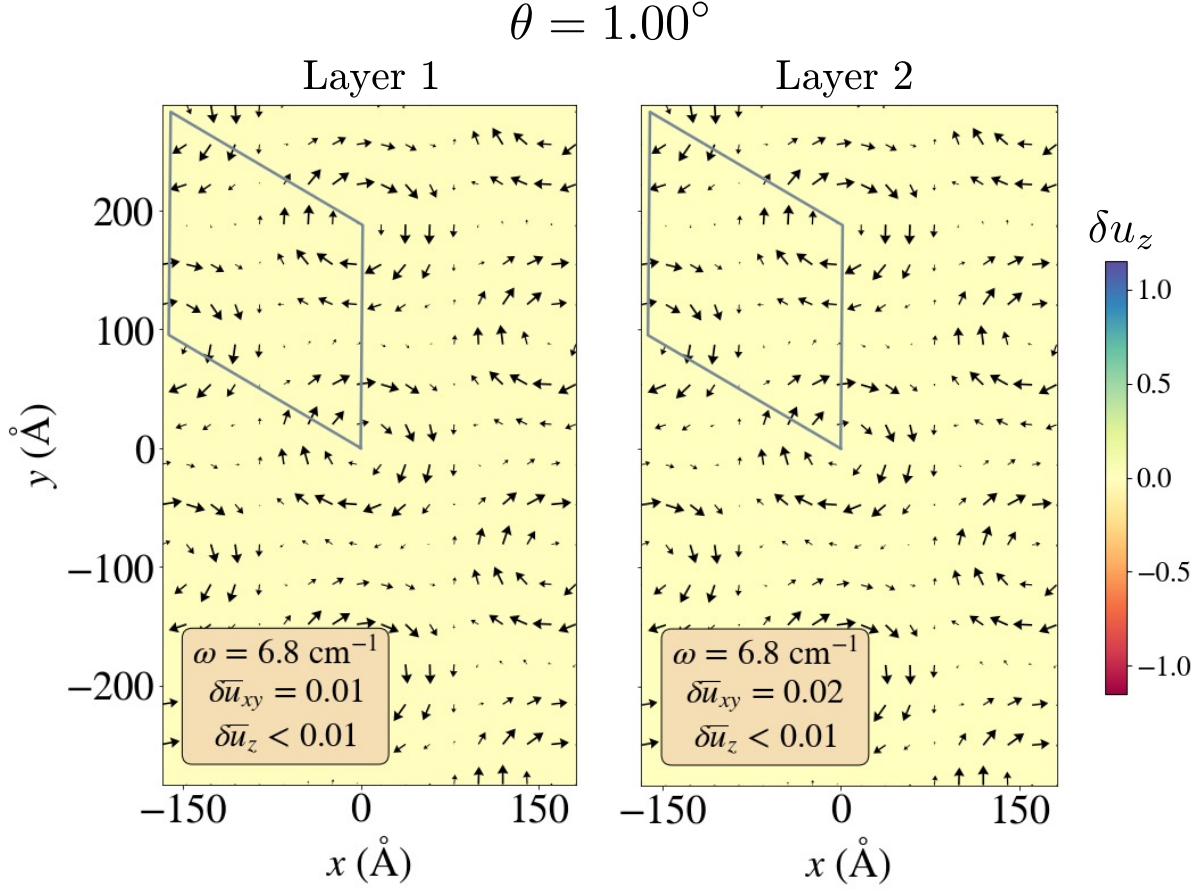}
    \caption{MoSe$_2$-WSe$_2$ SBLB mode real space atomic displacements at $\th = 1.00^\circ$. The insets and color scale are the same as in Fig.~\ref{fig:5-LB}.}
    \label{fig:8-SBLB}
\end{figure}

Finally, we examine the real space structure of SBLB mode in the MoSe$_2$-WSe$_2$ heterostructure. 
Unlike the LB mode where the in-plane motions are opposite at every real space position, the SBLB mode in the heterostructure breaks layer symmetry. As shown in Fig. \ref{fig:8-SBLB}, both layers share the same in-plane rotation direction.
Note that there is a preferential in-plane rotation direction in the heterostructure; we did not observe a degenerate SBLB mode with the other rotation direction. 
The out-of-plane motions are small but opposite between layers. 

The atomic displacements in real space may be examined beyond the $\G$ point, though that is beyond the scope of discussion in this paper. We provide a brief discussion in Appendix~\ref{app:not-gamma}.

\subsection{Higher energy LB (LB$_2$ and LB$_3$) Modes}\label{sec:results:subsec:HLB}
In addition to the lowest-energy LB mode analyzed above (which we now call LB$_1$), we observe two LB modes of higher frequency. 
The key characteristics of moir\'e LB modes are the oppositely oriented swirling in-plane motion and the opposite out-of-plane motion between each layer at every point $\vec{r}$; these are properties that all three LB modes exhibit. 
However, in LB$_1$, as shown in Fig.~\ref{fig:5-LB}, the out-of-plane displacement has a uniform direction within each layer. 
This does not hold for the LB$_2$ and LB$_3$ modes for all $\th$. 
To quantify the uniformity in out-of-plane displacement direction, we define the following quantity:
\begin{equation}\label{eqn:uniformity}
    \mathcal{U}_{\text{LB}}^{12} = \frac{1}{N^2} \sum_{i=1}^{N^2} \d_{\operatorname{sgn}[\d \vec{u}_z^{(1)}(\vec{r}_i)],1} \d_{\operatorname{sgn}[\d \vec{u}_z^{(2)}(\vec{r}_i)],-1},
\end{equation}
where $N$ is the sampling grid size in the moir\'e supercell and $\d \vec{u}_z^{(\ell)}(\vec{r}_i)$ is the $z$ component of the phonon at point $\vec{r}_i$ in layer $\ell$. 
Equation~\eqref{eqn:uniformity} essentially estimates the fraction of area where $\delta u_z^{(1)}$ points in the positive direction while  $\delta u_z^{(2)}$ points in the negative direction. 
Since it does not matter which layer breathes in which out-of-plane direction, we define the breathing uniformity value $\mathcal{U}_{\text{LB}} = \max(\mathcal{U}_{\text{LB}}^{12}, \mathcal{U}_{\text{LB}}^{21})$, which estimates the fraction of the moir\'e cell that breathes in the same out-of-plane direction. 
A translational mode with $z$ invariance has $\mathcal{U}_{\text{LB}} = 0$,
and a perfectly uniform breathing motion (such as that of LB$_1$) has $\mathcal{U}_{\text{LB}} = 1$. A maximally nonuniform breathing motion has $\mathcal{U}_{\text{LB}} = 0.5$, which means that at each point, the field is equally likely to point in either direction.  

\begin{figure*}[ht!]
    \centering
    \includegraphics[width=\textwidth]{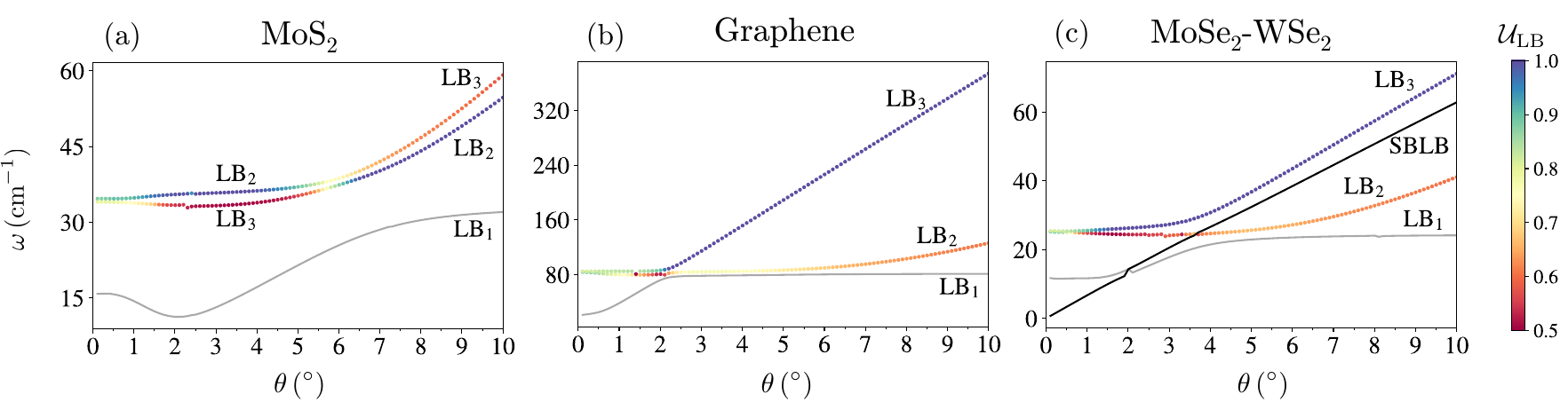}
    \caption{The frequencies of the LB$_2$ and LB$_3$ modes color-coded by their uniformity value $\mathcal{U}_{\text{LB}}$ (common scale on the right) for (a) MoS$_2$, (b) graphene, and (c) MoSe$_2$-WSe$_2$. LB$_1$ in grey and SBLB in black are shown for comparison.}
    \label{fig:5-DLB}
\end{figure*}

Figure~\ref{fig:5-DLB} shows the two higher-frequency LB modes for each material, color-coded by $\mathcal{U}_{\text{LB}}$. We observe that the uniformity of the breathing motion is dependent of $\theta$. In fact, at any value of $\th$, one mode is maximally uniform while the other is not. Hence, we define the LB$_2$ mode as the maximally uniform mode and the LB$_3$ mode as the other mode. This identification is not well-defined in degenerate transition regions, such as that near $\th = 6.0^\circ$ in MoS$_2$ and $\th \lesssim 1.0^\circ$ in all three materials. In MoS$_2$ only, the LB$_2$ and LB$_3$ mode change order in the frequency scale.

\begin{figure}[ht!]
    \centering
    \includegraphics[width=3.5in]{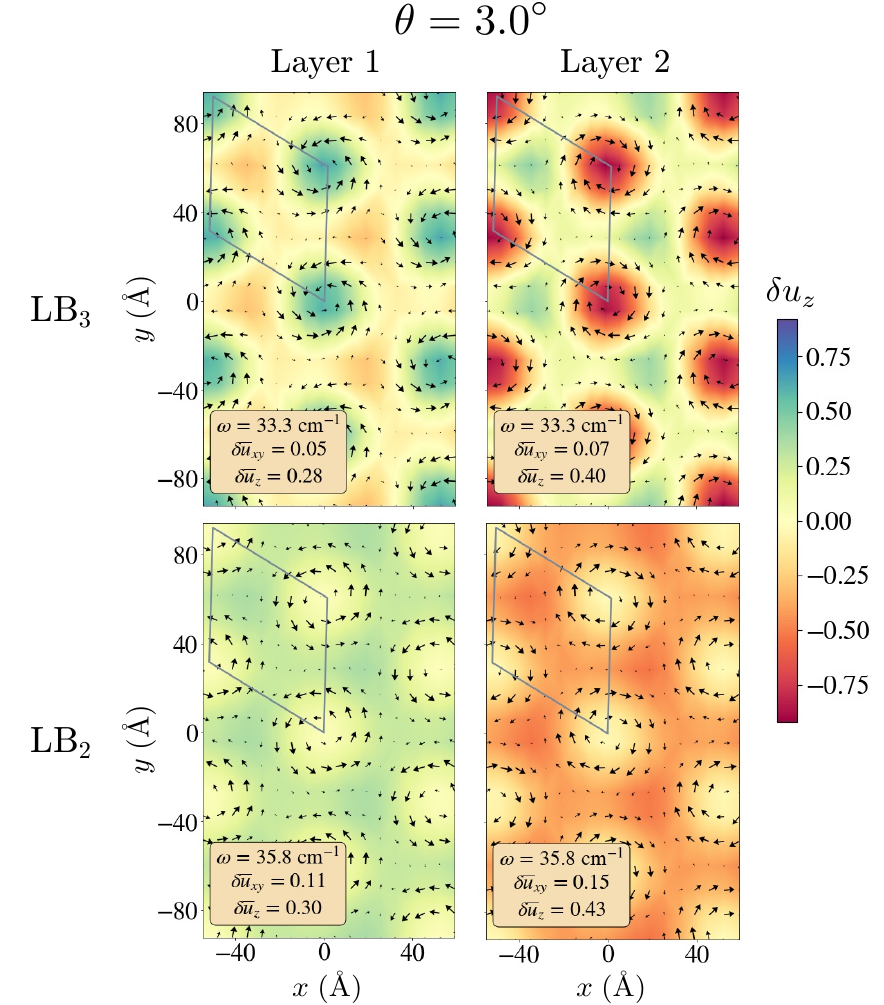}
    \caption{Atomic displacement fields of LB$_2$ and LB$_3$ in MoS$_2$ at $\theta=3.0^\circ$. The insets and color scale are the same as in Fig.~\ref{fig:7-shear}.}
    \label{fig:9-transition}
\end{figure}

In Fig.~\ref{fig:9-transition}, we provide an example of a uniform versus a non-uniform higher-frequency LB mode in MoS$_2$ at $\th = 3.0^\circ$. The non-uniformity manifests in the difference in direction of out-of-plane motion near the AA regions versus the AB/BA regions.

\section{Summary} \label{sec:summary}
We developed an accurate and efficient model for calculating phonon properties in bilayer moir\'e vdW heterostructures, which is based on the configuration space continuum (CSC) formalism. Within this formalism, we showed that phonons 
related to moir\'e patterns in twisted bilayers can be calculated directly from ensembles of phonons related to rigidly shifted bilayers at various arrangementss, with first-principles accuracy. 

By analyzing the low-frequency shearing (S) and layer breathing (LB) modes in three representative materials (twisted bilayer graphene, bilayer MoS$_2$, and a MoSe$_2$-WSe$_2$ heterostructure), 
we showed that the physics of moir\'e-pattern phonons depend on the twist angle $\th$.
Phonon frequencies vary slowly and smoothly with $\th$, while the atomic displacement magnitude is more sensitive to small changes in twist angle near certain $\th$ values.
Our CSC model can be applied more generally to classes of bilayers beyond the three materials on which we focus here, such as bilayers that are oppositely oriented (the pristine bilayer has one layer rotated by $180^\circ$ relative to the other) or Janus materials.

Despite its efficiency and accuracy, the CSC model is limited in its present form to lowest-frequency modes and negative shearing frequencies due to the approximations inherent in the model. 
In the work presented here, we only expand the moir\'e dynamical matrix to the first shell of the moir\'e Brillouin zone. Increasing the cutoff radius would allow for the extension to higher frequency modes, although a new derivation of the acoustic sum rule is required. 
Future work to broaden the scope of the CSC model may include introducing $\widetilde{\vec{k}}$ dependence on the off-diagonal components of the moir\'e dynamical matrix in order to remove the low-frequency approximation.
Moreover, the CSC model may be applied to further investigate relatively low-energy phonon modes of interest, such as chiral phonons~\cite{zhang2015chiral,suri2021chiral,maity2022chiral}.

Building upon previous experimental studies with Raman spectroscopy~\cite{campos2013raman,wu2014resonant,huang2016low,lin2018moire,quan2021phonon}, the phonon features derived by our CSC model may be used to guide further experimental comparison for other materials.
The most direct application of the CSC model is as a phonon framework for a model of electron-phonon coupling in moir\'e materials. 
Ultimately, such studies will provide a theoretical foundation for the origin of strongly correlated states in moir\'e systems, and the extent to which phonons play a role in such phenomena.

\begin{acknowledgments}
We thank Lukas Linhart, Florian Libisch, and Stephen Carr for their helpful advice. The calculations described in this paper were performed on the FASRC Cannon cluster supported by the FAS Division of Science Research Computing Group at Harvard University.
JL acknowledges funding support from the Harvard Herchel-Smith and PRISE fellowships. ZZ, DTL, and EK acknowledge funding from the STC Center for Integrated Quantum Materials, NSF Grant No. DMR-1231319; NSF DMREF Award No. 1922172; and the Army Research Office under Cooperative Agreement Number W911NF-21-2-0147.
ZZ and EK also acknowledge funding from the Simons Foundation, Award No. 896626.
DTL also acknowledges funding from the U.S. Department of Energy, Office of Science, under Award Number DE-SC0019300. MA acknowledges funding from the NSF under Award No. DMR-1922172
and the Army Research Office under Grant Number W911NF-14-1-0247.
\end{acknowledgments}

\newcommand{\intra} {\Phi^{\mathrm{intra}} (\bm{u})}
\renewcommand{\pdv}[2]{ \frac{\partial {#1}}{\partial {#2}}  }
\newcommand{\pddv}[2]{ \frac{\partial^2 {#1}}{\partial {#2}^2}  }
\newcommand{\pddvv}[3]{ \frac{\partial^2 {#1}}{\partial {#2} \partial {#3} }  }
\newcommand{\gradx}{\nabla_{ \bm{r}}}
\newcommand{\rr}[1]{\mathrm{#1}}
\begin{appendices}

\section{Generalized Stacking Fault Energy (GSFE)} \label{sec:GSFE}
\begin{figure}[ht!]
    \centering
    \includegraphics[scale=0.67]{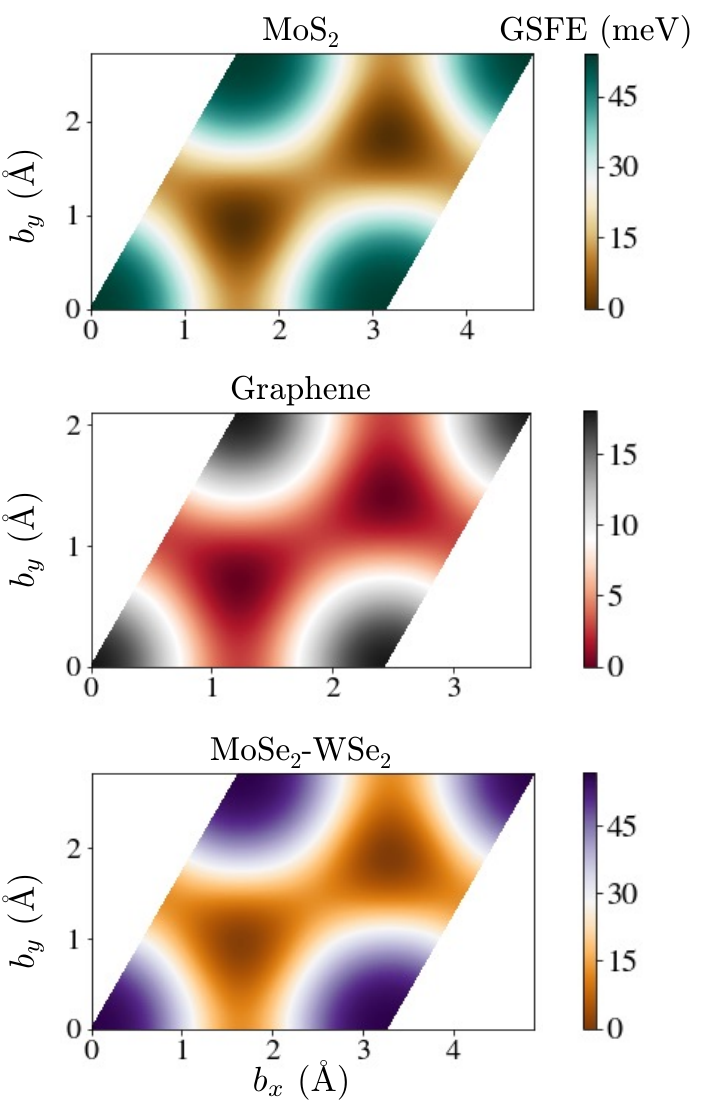}
    \caption{GSFE of MoS$_2$, graphene, and MoSe$_2$-WSe$_2$ fitted with first-order Fourier basis regression.}
    \label{fig:A-GSFE}
\end{figure}

The GSFE functional measures the energy as a function of stacking configuration. 
We evaluate the GSFE landscape $V^{\text{GSFE}}(\vec{b})$ according to a first shell expansion from \citet{carr2018relaxation}. Since the Fourier expansion is written in coefficients of $e^{i \vec{G} \cdot \vec{b}}$, with $\vec{G}$ a reciprocal lattice vector of the bilayer at a configuration $\vec{b}$, denote for a Bravais lattice basis $\vec{a}_1 = a_0 (1, 0), \vec{a}_2 = a_0 (1/2, \sqrt{3}/2)$: \begin{align}
    \begin{pmatrix} \m \\ \n \end{pmatrix} = \frac{2\pi}{a} \twomatrix{1}{-1/\sqrt{3}}{0}{2/\sqrt{3}} \vec{b}.
\end{align}
Using $120^\circ$ rotational and $(\m,\n) \to (\n,\m)$ symmetries, the GSFE can then be expanded to first shell as \begin{align}
\begin{aligned}
    V^{\text{GSFE}} = &\, c_0 + c_1 [\cos(\m) + \cos(\n) + \cos(\m+\n)] \\
    & + c_2 [\cos(\m+2\n) + \cos(\m-\n) + \cos(2\m+\n)] \\
    & + c_3 [\cos(2\m) + \cos(2\n) + \cos(2\m+2\n)] \\
    & + c_4 [\sin(\m) + \sin(\n) - \sin(\m+\n)]  \\
    & + c_5 [\sin(2\m + 2\n) - \sin(2\m) - \sin(2\n)]
\end{aligned}\label{eq:gsfe}
\end{align}
with $\mathbf{c} = (c_0, \cdots, c_5) \in \R^6$. We fit $V^{\text{GSFE}}$ with least-squares linear regression in the above basis. For quantities with inversion reciprocal-space symmetry, or equivalently $V^{\text{GSFE}}(\m,\n) = V^{\text{GSFE}}(-\m,-\n)$, $c_4 = c_5 = 0$. The parameters in $\mathbf{c}$ are important for computing the relaxation in configuration space \cite{carr2018relaxation,triGr}. Figure~\ref{fig:A-GSFE} shows the fits to the GSFE for each of the 3 bilayers we studied.
The coefficients we obtain for each material are given in Table~\ref{tab:GSFE}. 

\begin{table}
    \centering
    \begin{tabular}{|c|c|c|c|}
        \hline & MoS$_2$ & Graphene & MoSe$_2$-WSe$_2$ \\ \hline\hline
        $c_0$  & $25.91$ & $6.985$ & $27.07$ \\
        $c_1$  & $12.79$ & $4.111$ & $13.31$ \\
        $c_2$  & $-2.339$ & $-0.3104$ & $-2.335$ \\
        $c_3$  & $-8.0262$ & $-0.1023$ & $-0.8243$ \\
        $c_4$  & $0$ & $0$ & $0.1312$ \\
        $c_5$  & $0$ & $0$ & $0.2338$ \\ \hline
    \end{tabular}
    \caption{GSFE coefficients of MoS$_2$, graphene, and MoSe$_2$-WSe$_2$ heterostructure. All in units of meV per unit cell.}
    \label{tab:GSFE}
\end{table}

\section{\label{sec:details}Computational Details}
We study the moir\'e phonons of graphene, MoS$_2$, and the MoSe$_2$-WSe$_2$ heterostructure, with lattice constants $2.457,\, 3.178,\, 3.306$ \AA, respectively. Following the implementation in Fig. \ref{fig:workflow}, we sample a $9 \times 9$ uniform mesh in configuration space. For each configuration, we perform DFT calculations using the Vienna {\it{Ab initio}} Simulation Package (\texttt{VASP})~\cite{vasp1,vasp2,vasp3,vasp4} with the r$^2$SCAN-rVV10 van der Waals functional~\cite{r2scan}, an 800 eV plane-wave energy cutoff, $17 \times 17$ $k$-point grid, and $10^{-6}$ eV electronic convergence threshold. The horizontal positions of the carbon atoms (graphene) and metal atoms (TMDCs) are held fixed, but the remaining ionic coordinates are allowed to relax until the forces are below $10^{-6}$ eV/\AA. We then compute the interatomic force constants using the frozen phonon approach with a $3 \times 3$ supercell, converging electronic self-consistent loops to $10^{-8}$ eV.
We repeat the frozen phonon calculations for the monolayers, but use a $6 \times 6$ supercell for improved accuracy.

We then construct the rigid moir\'e dynamical matrix in the manner described in Section \ref{sec:csc:sub:cfg} of the main text, enforcing pristine monolayer and bilayer symmetries numerically with the \texttt{phonopy} and \texttt{hiPhive} libraries as well as the acoustic sum rule from Appendix \ref{app:sum_rule} \cite{phonopy,hiphive}. Finally, continuum relaxation is implemented by the procedure outlined in Section \ref{sec:csc:sub:reldm} from \citet{carr2018relaxation}. Continuum relaxation requires input parameters of the GSFE coefficients from Table \ref{tab:GSFE} and the monolayer bulk ($\mathcal{K}$) and shear ($\mathcal{G}$) moduli, given in Table \ref{tab:elastic}.

\begin{table}
    \centering
    \begin{tabular}{|c|c|c|c|c|}
    \hline
        & MoS$_2$ & Graphene & MoSe$_2$ & WSe$_2$ \\ \hline\hline
        $\mathcal{K}$ & 73546 & 99265 & 81194 & 73745 \\
        $\mathcal{G}$ & 59028 & 99275 & 71562 & 63681 \\ \hline
    \end{tabular}
    \caption{Elastic constants $\mathcal{K}$ (bulk modulus) and $\mathcal{G}$ (shear modulus) in units of meV per unit cell.}
    \label{tab:elastic}
\end{table}

Direct DFT calculations of moir\'{e} phonons were performed for a commensurate 7.34$^\circ$ graphene supercell containing 244 carbon atoms constructed from a primitive cell with lattice constant $a=2.467$ \AA. These calculations used \texttt{VASP} and the PBE exchange correlation functional~\cite{perdew1996generalized} with zero damping DFT-D3 van der Waals corrections~\cite{grimme2010consistent}, $3\times3\times1$ $k$-point sampling, and a plane wave energy cutoff of 400 eV. The atomic displacements, force constants, and phonon band structure were computed with the \texttt{phonopy} package. 

MD calculations of the moirè phonons are
performed using the LAMMPS package \cite{LAMMPS}.
Phonons are computed by direct diagonalization of the system's dynamical matrix \cite{Fabrizio_PRX2019} obtained after relaxing the twisted bilayer's atomic positions \cite{Fabrizio_PRB2018}.
The carbon-carbon intralayer interactions are modelled 
via the second generation REBO potential \cite{Brenner-JPhysCondMat2002}.
The interlayer interactions are instead modelled 
via the Kolmogorov-Crespi (KC) potential \cite{Kolmogorov-PPRB2005},
using the parametrization of Ref.~\cite{Ouyang-NanoLett2018}. 
The starting intralayer carbon-carbon distance is set equal to $a_0=1.3978$ \AA\,, 
corresponding to the equilibrium bond length of the adopted REBO potential, 
giving a lattice parameter of $a\approx2.42$ \AA.
Geometric optimizations are performed using the FIRE algorithm \cite{Bitzek-PRL2006}.
The atomic positions are relaxed toward equilibrium until the total forces acting on 
each atom become less than $10^{-6}$ eV/atom.

\section{Empirical Continuum Model} ~\label{sec:koshino}
One approach to obtaining moir\'e phonons is through an empirical continuum model based on the relaxation model introduced in~\citet{carr2018relaxation}. The approach is introduced by ~\citet{koshino2019moire}; we briefly review and adapt it here. 
To obtain the in-plane phonon modes, we perturb around the equilibrium positions after relaxation. 
Here, we assume a homobilayer, for which the relaxation in layers 1 and 2 have symmetry $\vecl{u}{1} (\vec{r}) = -\vecl{u}{2} (\vec{b}) \equiv \vec{u}.$
The equilibrium position is calculated by minimizing the total energy, which is the sum of the interlayer energy that is the integral of the GSFE in configuration space (Eq.~\eqref{eq:gsfe}),
\begin{align}
E^\rr{inter} (\vec{u}) = \int V^\rr{GSFE} (\vec{u}+\vec{b}) \, \dd^2 \vec{b},
\end{align}
and the intralayer elastic energy:
\begin{align}\label{eqn:intra_tot}
E^\rr{intra}(\vec{u}) &=
2\int \intra\,  \dd^2\vec{b},
\end{align}
with
\begin{align}
\intra = \frac{1}{2} \gradx  \vec{u} (\vec{b}):C:\gradx \vec{u} (\vec{b}),
\label{eqn:elasticity}
\end{align}
where $\gradx$ is the real space gradient and $C$ is the linear elasticity tensor.
For in-plane deformations, $C$ is a rank-4 tensor with its components defined as follows:
\begin{align}
	&C_{11ij} = \begin{pmatrix}
		\mathcal{K}+\mathcal{G} & 0 \\
		0 & \mathcal{G}
	\end{pmatrix} \quad
	C_{12ij} = \begin{pmatrix}
		0 & \mathcal{K}-\mathcal{G} \\
		\mathcal{G} & 0
	\end{pmatrix} \quad
	 \\
	&C_{21ij} = \begin{pmatrix}
		0 & \mathcal{G} \\
		\mathcal{K}-\mathcal{G} & 0
	\end{pmatrix} \quad
	C_{22ij} = \begin{pmatrix}
		\mathcal{G} & 0 \\
		0 & \mathcal{K}+\mathcal{G}
	\end{pmatrix}.\nonumber
\end{align}
A lattice vibration perturbs the equilibrium positions, leading to an additional kinetic energy contribution: 
\begin{equation}
T = \sum_{\ell=1}^2 \int \frac{\rho}{2} \left[\left(\dot{U}_x^{(\ell)}\right)^2 + \left(\dot{U}_y^{(\ell)}\right)^2 \right] \dd^2 \vec{r}, 
\end{equation} 
where $\rho$ is the area density of a single layer ($\rho=7.61\times10^{-7} \,\mathrm{kg/m^2}$ for graphene), $\ell$ labels the layer, and $\dot{U}^{(\ell)}_\alpha$ is the time derivative of the cartesian component $\alpha=x,y$ of $U$ in layer $\ell$. 

\begin{figure}[ht!]
  \centering
  \includegraphics[width=3.25in]{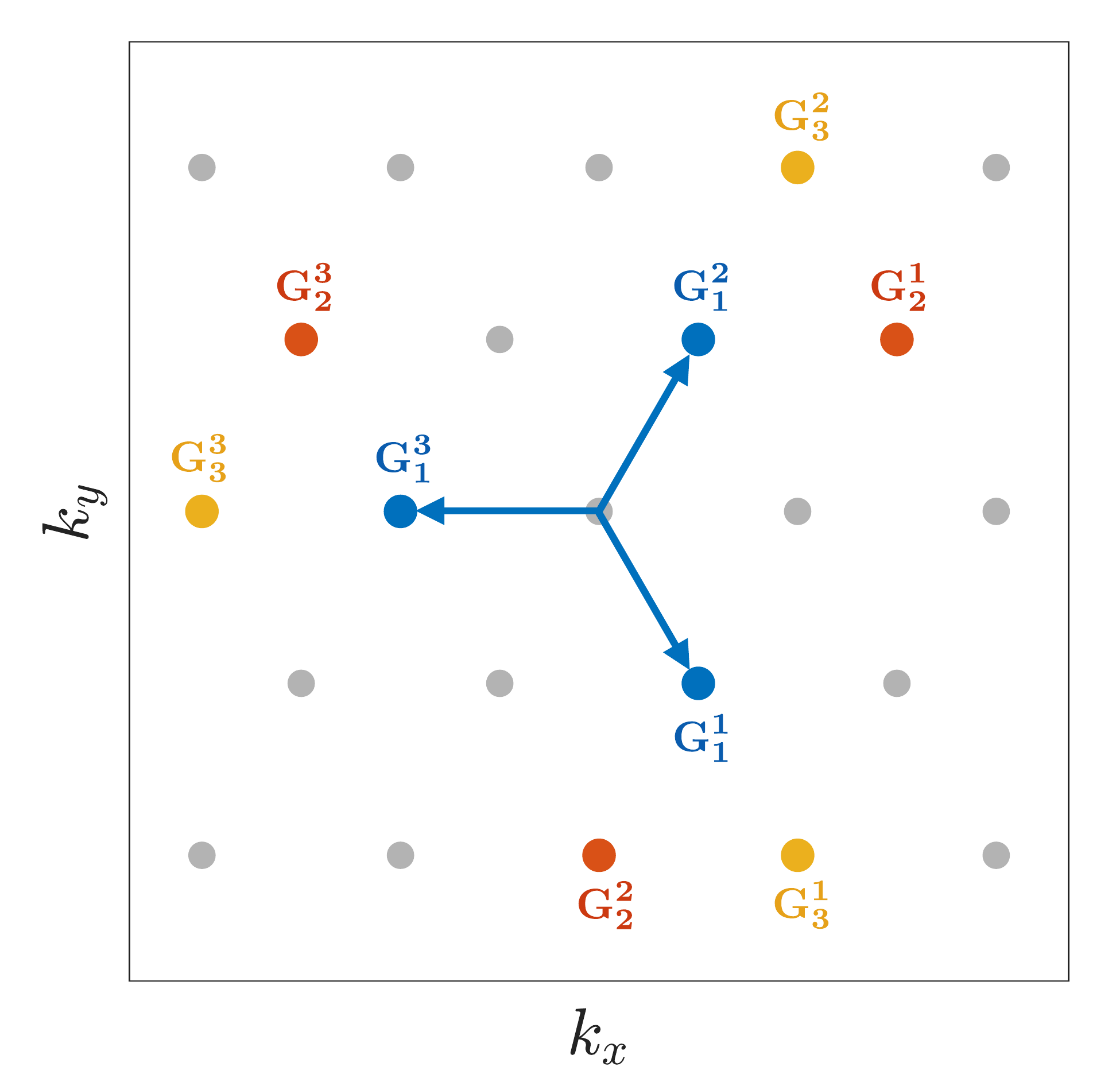}
  \caption[Definition of GSFE components]{Definition of reciprocal lattice vectors in Eq.~\eqref{eqn:gsfe_cos}. where the subscript labels different shells (different colors) and the superscript labels different components within the same shell (same color). The gray scatter points are reciprocal lattice vectors. }
  \label{fig:gvec}
\end{figure}

The Lagrangian of the system is given by $\mathcal{L}=T - (E^\mathrm{intra} + E^\mathrm{inter})$ as a function of $\vec{U}^{(\ell)} (\vec{r})$. We define $\vec{U}^{\pm} = \vec{U}^{(2)} \pm \vec{U}^{(1)}$ and rewrite $\mathcal{L}$ as a function of $\vec{U}^\pm$. 
Let us rewrite the GSFE from Eq.~\eqref{eq:gsfe} in a more compact form: 
\begin{align}
V^{\mathrm{GSFE}} (\vec{b}) = c_0 + \sum_{j=1}^3 \sum_{n=1}^3 c_j \cos(\vec{G}_j^n \cdot \vec{b}),\label{eqn:gsfe_cos}
\end{align} 
where $\vec{G}_j^n$ are the monolayer reciprocal lattice vectors defined in Fig.~\ref{fig:gvec}. Here, the subscript $j$ labels the shell (vectors that are the same distance from the origin) and the superscript $n$ labels the element within the shell. 
For simplicity we assume symmetry between AB and BA stacking and only consider the cosine components of the GSFE here ($c_1, c_2, c_3$), but the following results can be easily generalized to include the sine components ($c_4, c_5, c_6$). 

The equations of motion for $\vec{U}^-$ are given as follows:
\begin{widetext}
\begin{align}\label{eqn:eom}
\begin{aligned} 
\frac{1}{2} \rho \ddot{U}_x^- = \frac{1}{2} \left(\mathcal{K} + \frac{\mathcal{G}}{3} \right)\left(\frac{\partial^2 U_x^-}{\partial x^2} + \pddvv{U_y^-}{x}{y}\right) 
 + \frac{\mathcal{G}}{2} \left(\pddv{U_x^-}{x} + \pddv{U_x^-}{y} \right) +
 \sum_{j=1}^3 \sum_{n=1}^3 G_{j,x}^n c_j \sin(\vec{G}_j^n \cdot (\vec{b} + \vec{u^{-}})) \\
\frac{1}{2} \rho \ddot{U}_y^- = \frac{1}{2} \left(\mathcal{K} + \frac{\mathcal{G}}{3} \right)\left(\frac{\partial^2 U_y^-}{\partial y^2} + \pddvv{U_x^-}{x}{y} \right)
+ \frac{\mathcal{G}}{2} \left(\pddv{U_y^-}{x} + \pddv{U_y^-}{y} \right) 
+ \sum_{j=1}^3 \sum_{n=1}^3 G_{j,y}^n c_j \sin(\vec{G}_j^n \cdot (\vec{b} + \vec{u^{-}})),
\end{aligned}
\end{align}
\end{widetext}
where $G_{j,\alpha}^n $ is the $\alpha=x,y$ component of $\vec{G}_j^n$. 
Note that $\vec{U}(\vec{r})$ is the real space displacement and $\vec{u}(\vec{b})$ is the configuration space displacement and they are related by the linear transformation in Eq.~\eqref{eqn:mapping}. 
The terms on the 2nd lines of Eq.~\eqref{eqn:eom} that are proportional to $c_j$ come from the GSFE, which depends only on $\vec{U}^-$, the difference between relaxation in layer 1 and layer 2, and not on $\vec{U}^+$.
Thus the equations of motion for $\vec{U}^+$ are given by replacing $\vec{U}^-$ with $\vec{U}^+$ and dropping the terms proportional to $c_j$.

Because $\vec{U}^+$ represents equal motion in both layers, its fluctuations yield the acoustic phonons of a monolayer. 

To determine the moir\'e phonons we study $\vec{U}^-$ and consider a small vibration around the static equilibrium relaxation: 
\begin{align}
\vec{U}^- (\vec{r}, t) = \vec{U}_0^-(\vec{r}) + \delta \vec{U}^- (\vec{r}, t), 
\end{align}
where $\vec{U}_0^-$ is the static relaxation that minimizes the total energy, $E^\mathrm{intra} + E^\mathrm{inter}$, and $\delta \vec{U}^- (\vec{r}, t)$ is the perturbation around $\vec{U}_0^-$, which constitutes the phonon modes. 
We define the following Fourier coefficients to express the displacement and equations of motion in Fourier space: 
\begin{align}
\vec{U}_0^- (\vec{r}) = \sum_{\vec{G}} \vec{U}_{0,\widetilde{ \vec{G}}} e^{i\widetilde{\vec{G}} \cdot \vec{r}}, \nonumber \\
\delta \vec{U}^-(\vec{r}, t) = e^{-i\omega t} \sum_{\vec{q}}\delta\vec{U}^-_{\vec{q}} e^{i \vec{q}\cdot \vec{r}}, \nonumber \\
\sin (\vec{G}_j^n \cdot (\vec{b} + \vec{u^{-}})) = \sum_{\widetilde{\vec{G}}} f_{j,\vec{G}}^{n} e^{i \widetilde{\vec{G}} \cdot \vec{r}}, \nonumber \\
\cos (\vec{G}_j^n \cdot (\vec{b} + \vec{u^{-}})) = \sum_{\widetilde{\vec{G}}} h_{j,\vec{G}}^{n} e^{i \widetilde{\vec{G}}\cdot \vec{r}}, 
\end{align} 
where $\widetilde{\vec{G}}$ is the moir\'e reciprocal lattice vector. Note that $\widetilde{\vec{G}}\cdot \vec{r} = \vec{G}\cdot \vec{b}$, where $\vec{G}$ is the monolayer reciprocal lattice vector. We expand Eq.~\eqref{eqn:eom} around the static solution $\vec{U}_0$ to the first order in $\delta \vec{U}$ and perform a Fourier expansion to obtain the equation of motion for the perturbation part: 
\begin{equation} \label{eqn:koshino_dm}
\rho \omega^2 \delta \vec{U}^-_{\widetilde{\vec{G}} + \vec{q}} = \widehat{K}_{\widetilde{\vec{G}} + \vec{q}} \delta 
\vec{U}^-_{\widetilde{\vec{G}} + \vec{q}} 
- 4 \sum_{\widetilde{\vec{G}'}} \widehat{V}_{\widetilde{\vec{G}}-\widetilde{\vec{G}}'} \delta \vec{U}^-_{\widetilde{\vec{G}'} + \vec{q}},
\end{equation} 
where 
\begin{equation}
\widehat{K}_{\widetilde{\vec{G}}} = \begin{pmatrix} 
(\mathcal{K} + \frac{4}{3}\mathcal{G}) G_x^2 + \mathcal{G} G_y^2 & (\mathcal{K} + \frac{\mathcal{G}}{3}) G_x G_y \\ 
(\mathcal{K} + \frac{\mathcal{G}}{3}) G_x G_y & (\mathcal{K} + \frac{4}{3}\mathcal{G}) G_y^2 + \mathcal{G} G_x^2 
\end{pmatrix}
\end{equation} 
and 
\begin{equation}
\widehat{V}_{\widetilde{\vec{G}}} =\sum_{j=1}^3 c_j \sum_{n=1}^3
h_{j,\widetilde{\vec{G}}}^{n}
\begin{pmatrix} 
G_{j,x}^{n} G_{j,x}^{n} & G_{j,x}^{n} G_{j,y}^{n} \\
G_{j,x}^{n} G_{j,y}^{n} & G_{j,y}^{n} G_{j,y}^{n} 
\end{pmatrix}.
\end{equation} 
Here, the $\widehat{V}_{\widetilde{\vec{G}}-\widetilde{\vec{G}}'}$ is essentially an interlayer scattering selection rule that couples $\widetilde{\vec{G}}$ and $\widetilde{\vec{G}}'$
After obtaining the equilibrium relaxation displacement vectors $\vec{U}_0^-$ self-consistently~\cite{carr2018relaxation}, we solve the eigenvalue problem in Eq.~\eqref{eqn:koshino_dm} to obtain the eigenmodes $\omega^2$. 

\section{Further Discussion of the Soft Regime}\label{app:soft-regime}
A natural question from the analysis of the CSC model is whether the imaginary shearing frequencies in the soft regime are simply due to sampling issues. In this section we provide evidence that this is not the case.

\begin{figure}[ht!]
    \centering
    \includegraphics[width=\linewidth]{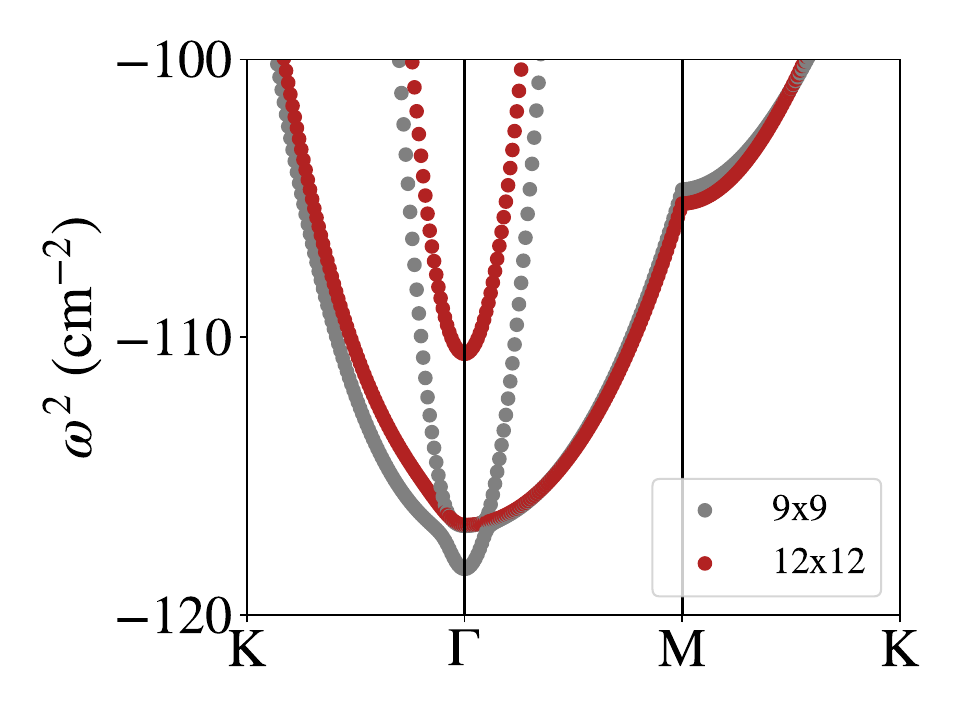}
    \caption{Eigenvalues $\w^2$ of the dynamical matrix of MoS$_2$ at $2.0^\circ$ for two configuration space sampling densities, $9\times9$ and $12\times12$.}
    \label{fig:dense}
\end{figure}

Our calculations in the continuum approach have used a $9 \times 9$ discretization of configuration space. Thus, one might imagine the possibility that this discretization was not sufficiently dense to cover enough configurations, and caused numerical error that led to soft modes. However, as Fig.~\ref{fig:dense} illustrates for bilayer MoS$_2$ with a representative twist angle of $\th = 2.0^\circ$, an improved sampling grid of $12 \times 12$ does not significantly increase the negative eigenvalues; in fact, if we were to plot the values of $\omega$, that is, the square root of the results shown in Fig.~\ref{fig:dense}, the difference is visually indistinguishable.

\begin{figure}[ht!]
    \centering
    \includegraphics[width=\linewidth]{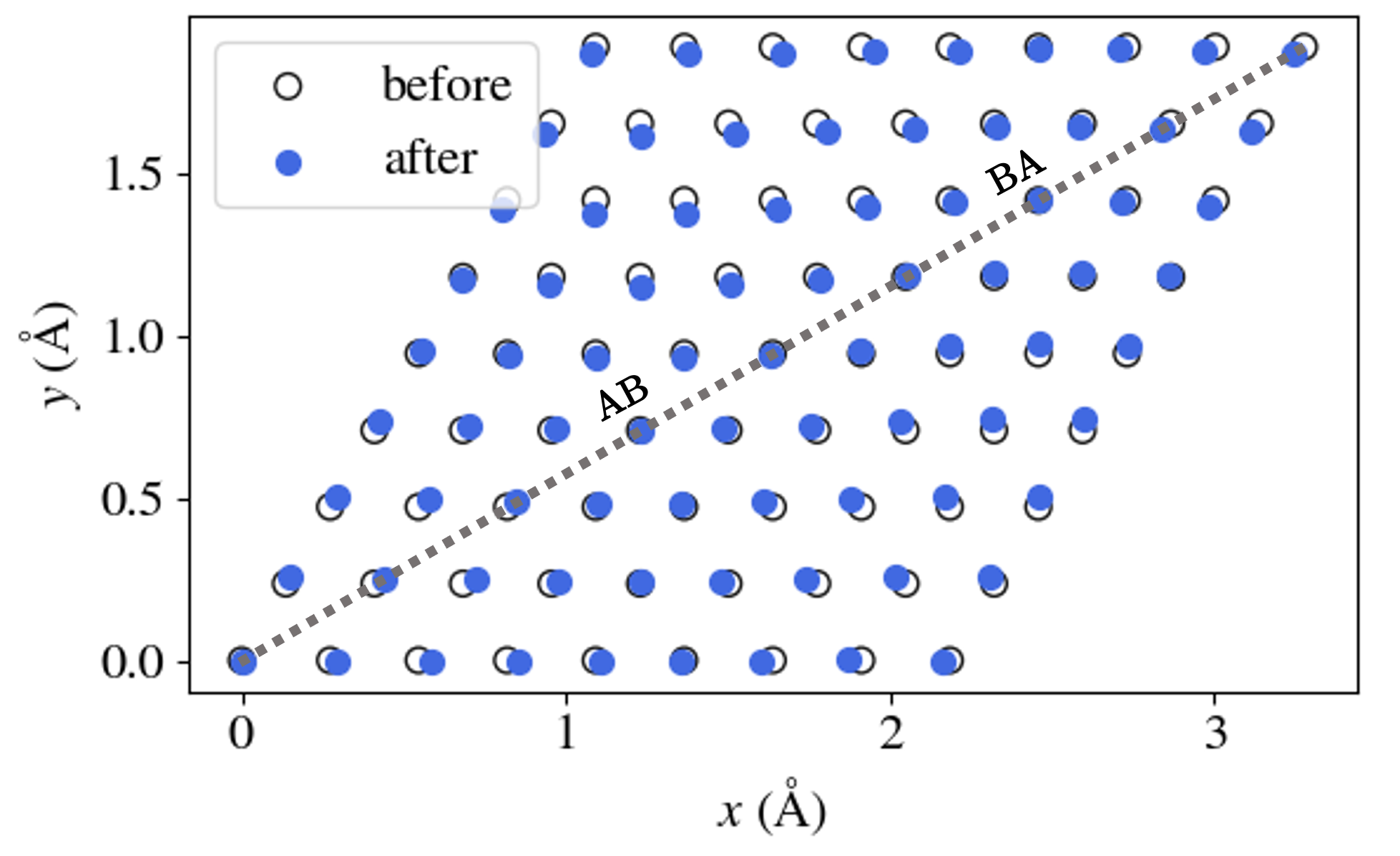}
    \caption{Updated configurations of MoS$_2$ at $\th = 2.0^\circ$ after continuum relaxation.}
    \label{fig:relaxBA}
\end{figure}

Moreover, as mentioned earlier, relaxation cannot alone fix the softness of the shearing modes. For a twist angle of $2.0^\circ$ the relaxation produces only very small changes to the configuration vectors $\vec{b}$, as shown in Fig.~\ref{fig:relaxBA}. Hence, the problem of ultrasoft shearing modes in the soft regime is a problem more subtle than that of sampling or relaxation in configuration space.

\section{Real Space Displacement at Other High-Symmetry Points}\label{app:not-gamma}

\begin{figure*}[ht!]
    \centering
    \includegraphics[width=\linewidth]{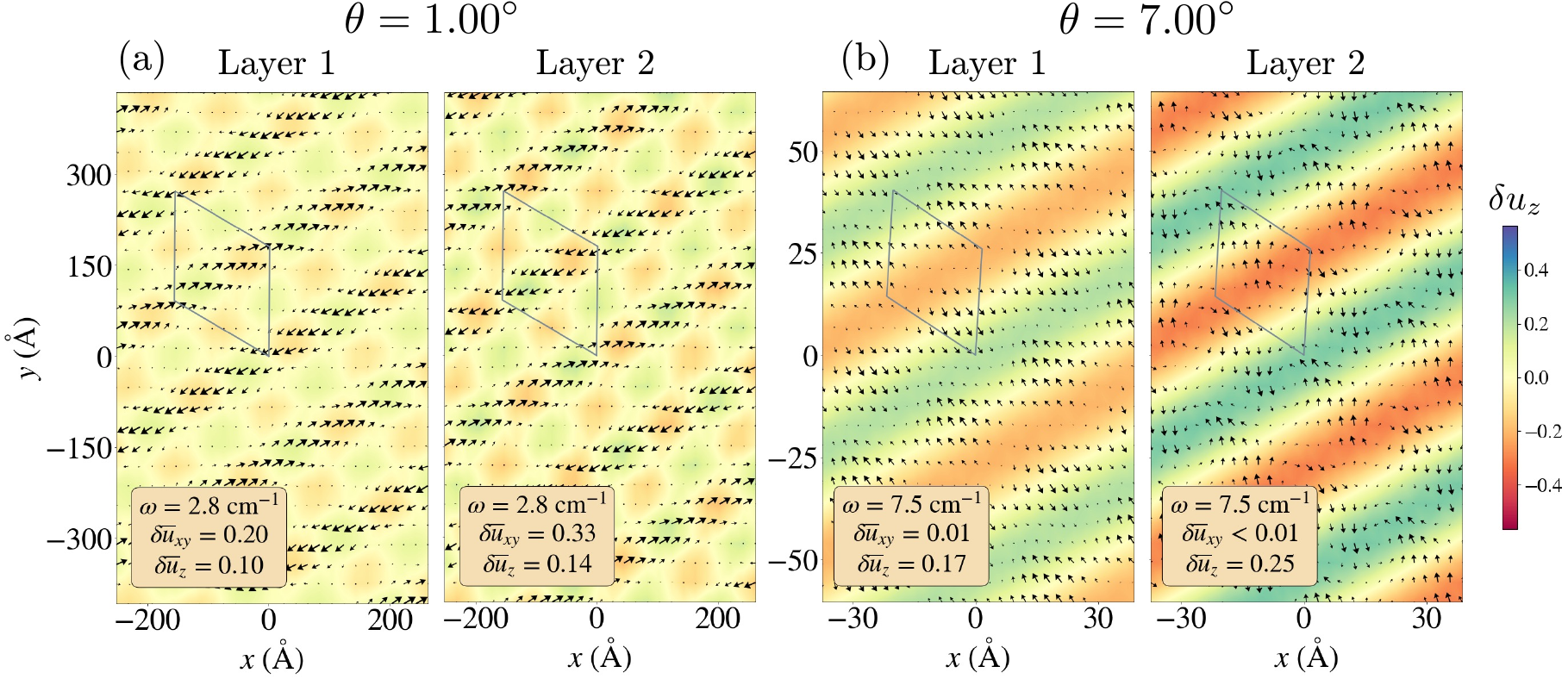}
    \caption{Displacement fields of the first folded band in MoS$_2$ at $M$. For each of (a) and (b), the left (right) side displays layer 1 (2). The periodicity is of a $2 \times 2$ moir\'e supercell.}
    \label{fig:A-M}
\end{figure*}

\begin{figure*}[ht!]
    \centering
    \includegraphics[width=\linewidth]{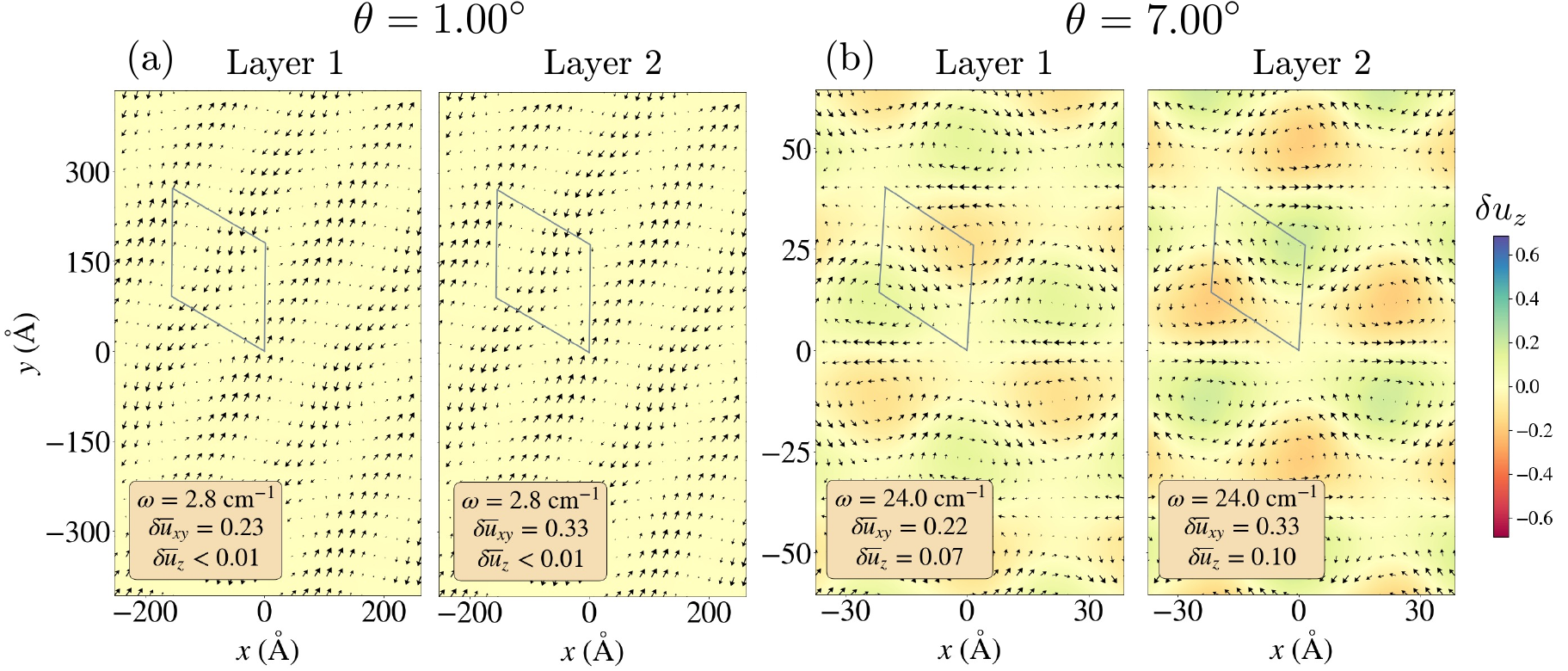}
    \caption{Displacement fields of the first folded band in MoS$_2$ at $K$. For each of (a) and (b), the left (right) side displays layer 1 (2). The maximum periodicity is of a $3 \times 3$ moir\'e supercell.}
    \label{fig:A-K}
\end{figure*}

The real space analysis in this paper has focused on the low-energy modes at $\G$. However, the CSC model permits analysis of fields at any $k$-point, computed in the same way but with an extra phase given by the $k$-point to the inverse Fourier transform. Every field at $\G$ has the period of one moir\'e cell $\l$. However, at $M$, some (but not all) bands have the period of 2 moir\'e lengths, while at $K$ some have the period of 3 moir\'e lengths. This is to be expected, as for a Bravais hexagonal lattice with basis vectors oriented $120^\circ$ apart, $M = \frac{1}{2} \widetilde{\vec{g}}_1 + \frac{1}{2} \widetilde{\vec{g}}_2$ and $K = \frac{2}{3} \widetilde{\vec{g}}_1 + \frac{1}{3} \widetilde{\vec{g}}_2$ where $\set{\widetilde{\vec{g}}_i}$ is the reciprocal lattice basis. Examples of these periodicities are shown respectively in Figs. \ref{fig:A-M} and \ref{fig:A-K} on MoS$_2$ on the first folded band, which is the lowest-energy band at each $k$-point stemming from the first folded cluster. For example, the analogous band for graphene is the cluster between the S and LB modes of Fig.~\ref{fig:4-bands}(a).

\end{appendices}

\bibliography{apssamp}

\end{document}


\preprint{APS/123-QED}

\title{Erratum: Low-energy moir\'e phonons in twisted bilayer van der Waals heterostructures [Phys. Rev. B 106, 144305 (2022)]}

\author{Jonathan Z. Lu}
\author{Ziyan Zhu}
\author{Mattia Angeli}
\author{Daniel T. Larson}
\author{Efthimios Kaxiras}

\date{\today}

\maketitle

Due to the many complicated phonon modes arising in moir\'{e} structures, the identification of those modes can be challenging.
Subsequent to the publication of our paper\cite{lu2022low}, we devised a more consistent method for classifying the low-energy phonons.  We emphasize that the numerical values of the phonon spectrum calculated using our Configuration Space Continuum (CSC) model are unchanged, and only the identification of the band character is modified.
In particular, the correct identification of the primary layer-breathing (LB) mode in Fig.~5 of Ref.\cite{lu2022low} has implications for the interpretation of several of the subsequent figures and accompanying text. Below we give an explanation of each affected figure.

\begin{figure*}[ht!]
    \centering
    \includegraphics[width=\linewidth]{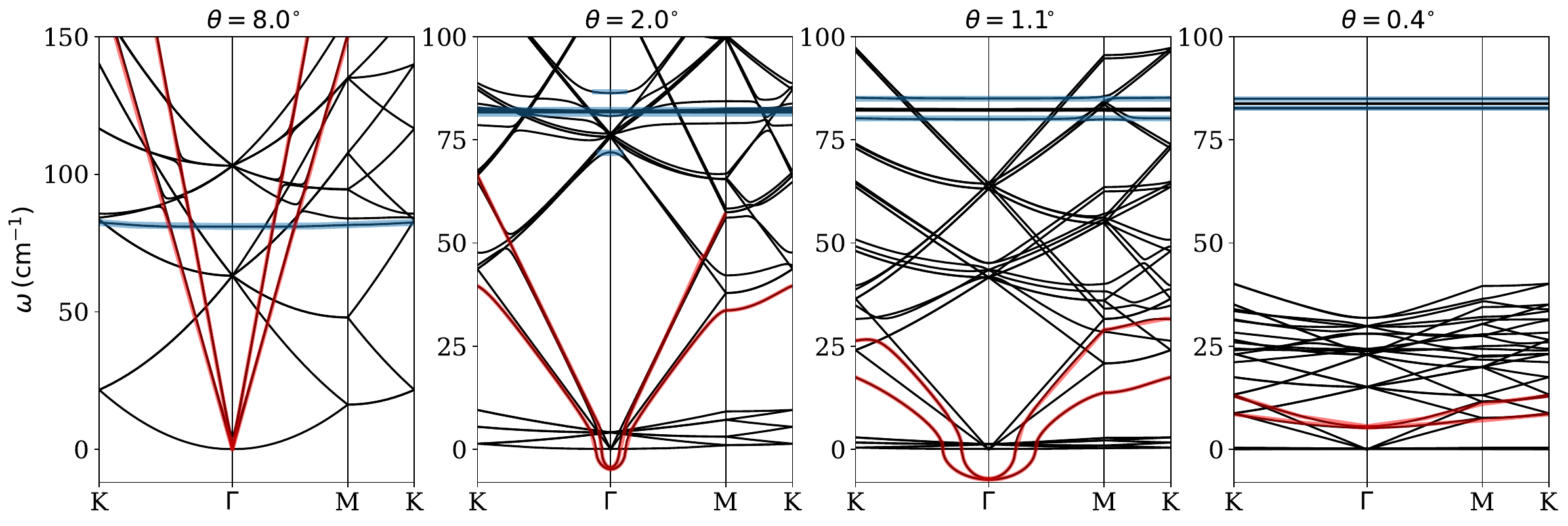}
    \caption{[Figure 5 in Ref.~\cite{lu2022low}] Phonon bands of graphene as obtained from our CSC model at four representative angles of (a) $8.0^\circ$, (b) $2.0^\circ$, (c) $1.1^\circ$, and (d) $0.4^\circ$. The two-layer shearing modes are highlighted in red, and the layer breathing (LB) modes in blue.}
    \label{fig:new-bands}
\end{figure*}

\textbf{Figure 5 of Ref.~\cite{lu2022low}} In Fig.~\ref{fig:new-bands} we re-plot the phonon spectrum for twisted bilayer graphene over a larger range of frequencies and make a more accurate identification of the layer breathing modes.
For $\theta=8.0^\circ$ the fundamental LB mode arising from the $\widetilde{\mathbf{G}}_0 = \mathbf{0}$ components of the moir\'{e} dynamical matrix crosses the $\Gamma$ point of the moir\'{e} Brillouin zone at 81.0 cm$^{-1}$. 
The group of six bands that cross $\Gamma$ at 103 cm$^{-1}$ are zone-folded breathing modes arising from the six $\widetilde{\mathbf{G}}_i$ components in the first shell of the moir\'{e} momentum expansion. 
As the twist angle decreases, the folded components of the LB mode, as well as the folded components of the acoustic and shearing modes, all decrease in frequency due to the reduction in the size of the moir\'{e} Brillouin zone. 
For $\theta=2.0^\circ$ the fundamental LB mode and the folded copies are all compressed into a narrow frequency range around 81 cm$^{-1}$ that is highlighted in blue. 
In addition, the hybridization of these modes with the folded copies of the transverse acoustic (TA) mode leads to lower and higher frequency modes that show significant out-of-plane breathing character. They have frequencies of 72 and 86 cm$^{-1}$ at $\Gamma$ and are also highlighted in blue.  
For smaller angles, the principal and folded LB modes hybridize producing lower and higher frequency breathing modes that have displacements only in the AA and AB/BA regions of the real space moir\'{e} unit cell, respectively. 
Due to the truncated expansion to the first-shell in the moir\'e reciprocal lattice vectors, there are missing higher frequency modes, most notably the frequency range above $40 \, \mathrm{cm}^{-1}$ for $\theta=0.4^\circ$. Since we are focusing on low-frequency phonons, we leave the higher-shell expansion to future work. 

\textbf{Figure 6 of Ref.~\cite{lu2022low}}
The blue line plotted in each of the three panels shows the frequency of the LB mode in the decoupled region (large angles, blue shade), and the frequency of an in-plane mode in the relaxed region (small angles, brown shade), which is why there is a significant drop in the blue curve for small angles. 
The frequencies of the LB modes across all angles are shown more clearly in Figure 11 of Ref.~\cite{lu2022low}. 
In that figure, for large angles the mode labeled LB$_1$ corresponds to the principal LB mode originating from the $\widetilde{\mathbf{G}}_0$ component of the moir\'{e} dynamical matrix and the mode labeled LB$_2$ corresponds to the set of folded LB modes originating from the $\widetilde{\mathbf{G}}_i$ components in the first shell of moir\'{e} reciprocal lattice vectors. 
For small angles, the modes LB$_2$ and LB$_3$ correspond to the hybridized breathing modes, highlighted in blue in Fig.~\ref{fig:new-bands} for graphene with $\theta=1.1^\circ$ and $0.4^\circ$, that exhibit vibration in either the AA or AB/BA regions, as shown in Figure 12 of Ref.~\cite{lu2022low} for an MoS$_2$ bilayer.

\textbf{Figure 7 of Ref.~\cite{lu2022low}}
The decrease in the $\widetilde{\mathbf{G}}_0$ contribution for small angles demonstrates the increasing contribution of zone-folded phonons to the modes whose frequencies are plotted in Figure 6 of Ref.~\cite{lu2022low}.

\textbf{Figure 9 of Ref.~\cite{lu2022low}}
The mode depicted in panel (a) has in-plane displacements larger than the out-of-plane displacements and represents an in-plane mode that shows an imprint of the moir\'{e} pattern in the form of the rotations about the AA regions at the corner of the real space moir\'{e} unit cell (grey parallelogram). The atomic displacements of the breathing modes in bilayer graphene for $\theta=1.00^\circ$ look qualitatively just like those in Figure 12 of Ref.~\cite{lu2022low}, with the lower frequency mode ($\omega=80.2$ cm$^{-1}$) showing significant out-of-plane displacement near the AA regions (corners of the moir\'{e} cell) and little displacement in the AB and BA regions between them. The higher frequency mode ($\omega=85.0$ cm$^{-1}$) has the reverse pattern: significant vertical displacement in the AB/BA regions and little displacement in the AA regions.

\textbf{Figure 10 of Ref.~\cite{lu2022low}}
This is a folded, in-plane acoustic mode that shows an imprint of the moir\'{e} pattern in the form of the rotations about the AA regions at the corner of the real space moir\'{e} unit cell (grey parallelogram).

\textbf{Figure 11 of Ref.~\cite{lu2022low}}
For large angles the mode labeled LB$_1$ corresponds to the principal LB mode originating from the $\widetilde{\mathbf{G}}_0$ component of the moir\'{e} dynamical matrix and the mode labeled LB$_2$ corresponds to the set of folded LB modes originating from the $\widetilde{\mathbf{G}}_i$ components in the first shell of moir\'{e} reciprocal lattice vectors. 
For small angles the modes LB$_2$ and LB$_3$ correspond to the hybridized breathing modes, highlighted in blue in Fig.~\ref{fig:new-bands} for graphene with $\theta=1.1^\circ$ and $0.4^\circ$, that exhibit vibrations in either the AA or AB/BA regions, as shown in Figure 12 of Ref.~\cite{lu2022low} for an MoS$_2$ bilayer.
The modes labeled LB$_3$ and SBLB represent linearly dispersing acoustic modes that are folded back onto $\Gamma$ and therefore have frequencies that decrease linearly with the twist angle due to the reduction in the size of the moir\'{e} Brillouin zone.

While this Erratum corrects the interpretation of the figures previously presented, the improved identification of the phonon modes has further implications for the detailed microscopic understanding of lattice vibrations within moir\'{e} structures that warrants more extensive discussion that will be presented in a future publication.

%